\documentclass[11pt,a4paper]{article}
\pdfoutput=1
\usepackage{docstyle}
\usepackage{float}
\usepackage{upgreek}
\usepackage{subcaption}
\usepackage{caption}
\usepackage{soul}

\usepackage{hyperref}

\usepackage[table-parse-only,per-mode=symbol-or-fraction,separate-uncertainty]{siunitx}
    \DeclareSIUnit\bar{bar}

\renewcommand\({\left(}
\renewcommand\){\right)}

\newcommand{\be}{\begin{equation}}
\newcommand{\ee}{\end{equation}}
\newcommand{\bea}{\begin{eqnarray}}
\newcommand{\eea}{\end{eqnarray}}

    \newcommand{\rchi}{\raisebox{\depth}{$\chi$}}
	
\begin{document}

\preprint{MPP-2023-136}

\title{A proposal for a low-frequency axion search in the 1--2 \textmu eV range and below with the BabyIAXO magnet}

\author [1]{S.~Ahyoune,}
\author [2]{A.~\'Alvarez~Melc\'on,}
\author [1]{S.~Arguedas~Cuendis,}
\author [3]{S.~Calatroni,}
\author [*,4]{C.~Cogollos,}
\author [5]{J.~Devlin,}
\author [2]{A.~D\'iaz-Morcillo,}
\author [6]{D.~D\'iez-Ib\'añez}
\author [4]{B.~D\"obrich,}
\author [7]{J.~Galindo,}
\author [8]{J.D. Gallego,}
\author [2]{J.M.~Garc\'ia-Barcel\'o,}
\author [9]{B.~Gimeno,}
\author [3,10]{J.~Golm,}
\author [6]{Y.~Gu,}
\author [4,11]{L.~Herwig,}
\author [6]{I.~G.~Irastorza,}
\author [2]{A.J.~Lozano-Guerrero,}
\author [12]{C.~Malbrunot,}
\author [1,13]{J.~Miralda-Escud\'e,}
\author [2]{J.~Monz\'o-Cabrera,}
\author [2]{P.~Navarro,}
\author [2]{J.R.~Navarro-Madrid,}
\author [6]{J.~Redondo,}
\author [9]{J.~Reina-Valero,}
\author [14]{K.~Schmieden,}
\author [14]{T.~Schneemann,}
\author [15]{M.~Siodlaczek,}
\author [16,17]{S.~Ulmer}
\author [3]{and W.~Wuensch}

\affiliation[1]{Institut de Ci\`encies del Cosmos, Universitat de Barcelona, 08028 Barcelona, Spain}
\affiliation[2]{Department of Information and Communications Technologies, Technical University of Cartagena, 30203 Cartagena, Spain}
\affiliation[3]{CERN - European Organization for Nuclear Research, Geneva, Switzerland}
\affiliation[4]{Max-Planck-Institut f\"{u}r Physik, F\"{o}hringer Ring 6, 80805 M\"{u}nchen, Germany}
\affiliation[5]{Imperial College London, Blackett Laboratory, Prince Consort Road, London SW7 2BW, U.K.}
\affiliation[6]{Center for Astroparticles and High Energy Physics (CAPA), Universidad de Zaragoza, 50009 Zaragoza, Spain}
\affiliation[7]{Instituto Tecnol\'ogico de Arag\'on, 50018 Zaragoza, Spain}
\affiliation[8]{Yebes Observatory (IGN), 19141 Guadalajara, Spain}
\affiliation[9]{Instituto de Física Corpuscular (IFIC), CSIC-University of Valencia, 46980 Valencia, Spain}
\affiliation[10]{Institute  for  Optics  and  Quantum  Electronics,  Friedrich  Schiller  University  Jena, 07743 Jena,  Germany}
\affiliation[11]{Technische Universit\"at M\"unchen,  Arcisstraße 21, 80333 M\"unchen, Germany}
\affiliation[12]{TRIUMF, 4004 Wesbrook Mall, Vancouver, BC V6T 2A3, Canada}
\affiliation[13]{Instituci\'o Catalana de Recerca i Estudis Avan\c cats, 08010 Barcelona, Spain}
\affiliation[14]{Johannes Gutenberg Universit\"{a}t Mainz, Staudingerweg 7, 55128 Mainz, Germany}
\affiliation[15]{Technical University of Darmstadt, Institute for Energy Systems and Technology, 64287 Darmstadt, Germany}
\affiliation[16]{RIKEN, Fundamental Symmetries Laboratory, Wako, Japan}
\affiliation[17]{Heinrich  Heine  University,  D\"{u}sseldorf,  Univerist\"{a}tsstrasse  1,  D-40225  D\"{u}sseldorf,  Germany}

\affiliation[*]{Corresponding author}
\emailAdd{cogollos@mpp.mpg.de}

\vspace{5cm}
\noindent

\abstract{
In the near future BabyIAXO will be the most powerful axion helioscope, relying on a custom-made magnet of two bores of 70 cm diameter and 10 m long, with a total available magnetic volume of more than 7 m$^3$.
In this document, we propose and describe the implementation of low-frequency axion haloscope setups suitable for operation inside the BabyIAXO magnet. The RADES proposal has a potential sensitivity to the axion-photon coupling $g_{a\gamma}$ down to values corresponding to the KSVZ model, in the (currently unexplored) mass range between 1 and 2 \textmu eV, after a total effective exposure of 440 days. This mass range is covered by the use of four differently dimensioned 5-meter-long cavities, equipped with a tuning mechanism based on inner turning plates. A setup like the one proposed would also allow an exploration of the same mass range for hidden photons coupled to photons. An additional complementary apparatus is proposed using LC circuits and exploring the low energy range ($\sim 10^{-4}-10^{-1}$ \textmu eV). The setup includes a cryostat and cooling system to cool down the BabyIAXO bore down to about 5 K, as well as appropriate low-noise signal amplification and detection chain.}

\maketitle

\section{Introduction}

Two of the most pressing issues of fundamental particle physics, the strong CP problem~\cite{Bigi:2000yz,Peccei:2006as,Dragos:2019oxn} and the identification of the nature of the Dark Matter (DM) of the Universe~\cite{Bertone:2016nfn}, could be solved by the existence of the so-called QCD axion~\cite{Weinberg:1977ma,Wilczek:1977pj}. The axion is a hypothetical spin 0 particle, with negative intrinsic parity, whose existence is a direct implication of the Peccei-Quinn mechanism to explain the strong CP problem, being the pseudo-Nambu-Goldstone boson of a new global, axial and spontaneously-broken U(1) symmetry, which suffers from the color anomaly~\cite{Peccei:1977hh,Peccei:1977ur}. 

Although there are several ideas for solving the strong CP problem and for the nature of DM, the axion hypothesis has the advantage of solving both problems and of being in principle discoverable with current technology. 
The main free parameter of the PQ mechanism, the axion decay constant $f_a$, sets the axion mass ($m_a$) and the typical values of the interaction strengths to standard model particles. Usually, the experimentally most sensitive coupling is the one to photons: $g_{a\gamma}\propto 1/f_a$.  Thus, it was not difficult to exclude the existence of relatively strongly coupled axions like the original PQ model, which connected $f_a$ with the electroweak scale ($f_a\sim O(m_W)$), see \cite{DiLuzio:2020wdo} for a review. Invisible axion models with $f_a \gg \rm GeV$ require much more sophisticated techniques to find the axion, but a healthy collection of techniques and ideas has been put forward in the last four decades, see~\cite{Irastorza:2018dyq}. 

The most mature techniques, axion haloscopes and helioscopes, were proposed long ago and have been remarkably developed since. The IAXO helioscope proposal will be able to look for QCD axions at the meV mass frontier~\cite{IAXO:2019mpb}, while cavity haloscopes like ADMX \cite{ADMX:2020ote}, HAYSTAC \cite{HAYSTAC:2023cam} and CAPP \cite{Yoon:2022gzp} are currently being operated with discovery sensitivity for axions in the galactic DM halo if they have masses of a few \textmu eV. 

\begin{figure}[htbp]
\begin{center}
\includegraphics[width=\textwidth]{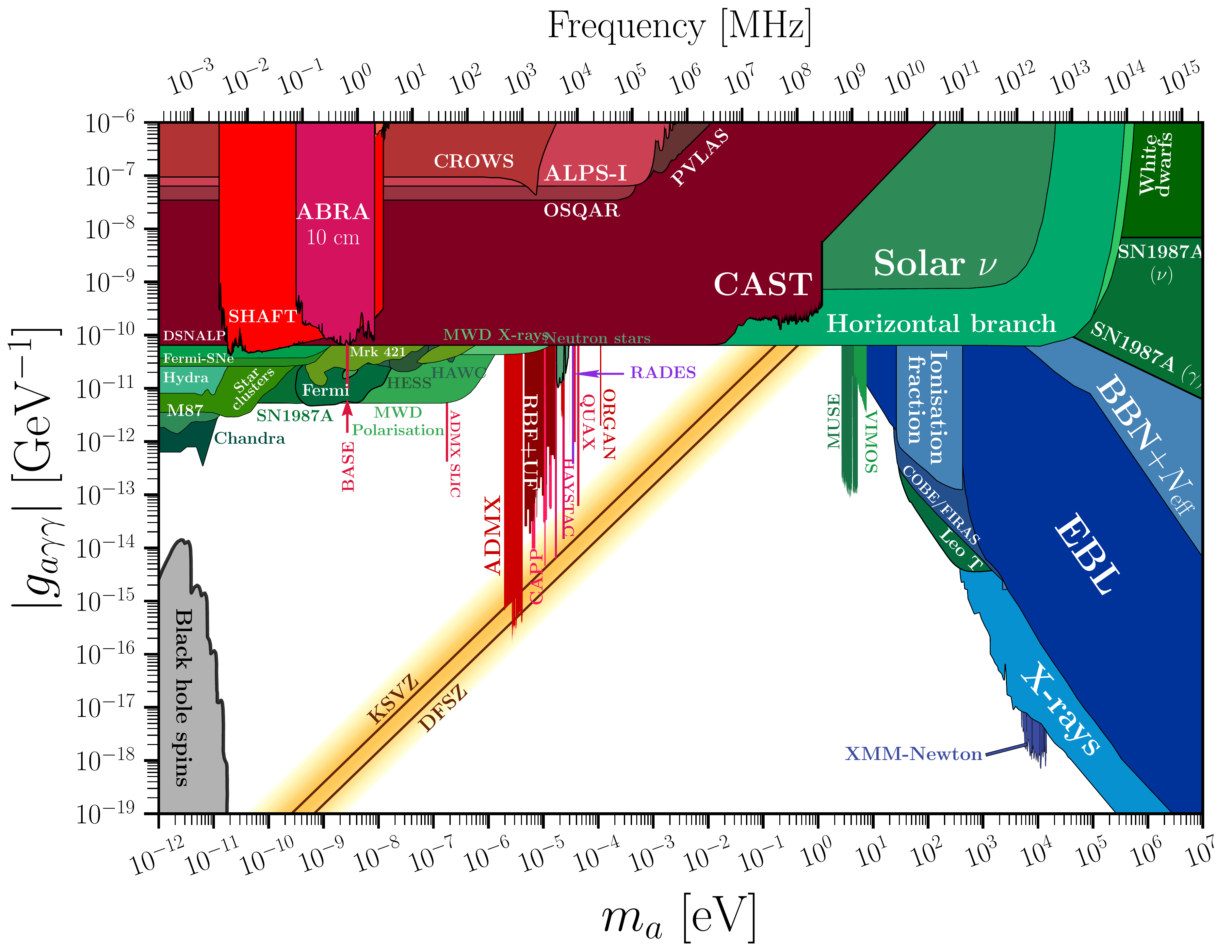}
\caption{Experimental exclusions for axion-like particles in axion-photon coupling vs. mass plane. From \cite{ciaran}.}
\label{gag}
\end{center}
\end{figure}

The search for galactic DM axions is extremely challenging due to the small interaction of axions and SM particles. The cavity haloscope of Sikivie employs the axion-2-photon vertex and a strong static B-field to convert axion DM into microwaves {\em resonantly}, but the ensuing enhancement (up to $Q_a\sim 10^6$) can only be achieved if the cavity is made to resonate at the Compton frequency of the axion, given by its mass. Since the mass is not known or specified by the PQ mechanism, it is unavoidable to scan over different cavity resonant frequencies until a signal is detected, or on the other hand upper limits can be set. Following the lead of the ADMX collaboration, several experiments are performing this scan. 
From Figure \ref{gag}, one can deduce that it is easier to build and run experiments in the few \textmu eV range. There is an inclination for new experiments to aim at higher-masses. The motivation for the latter stems from theoretical grounds that are worth reviewing. 

The axion DM yield from the big bang is calculable a priori if one specifies an axion model (with UV completion) and a detailed cosmological history. In practice, the most important difference between predictions is dictated by the presence or absence of a PQ symmetry restoration phase after cosmic inflation. One can then differentiate the pre and post-inflation scenarios. In the post-inflationary scenario, different causal patches of our observable Universe start with different uncorrelated initial conditions and the final DM yield is an average over possible initial conditions. 
In the pre-inflationary scenario, inflation selects one of those patches and only anthropic arguments can be invoked to reduce the uncertainty in initial conditions, which is maximal.  

The next most important difference is the expansion rate of the Universe at the time when the axion field starts responding to its potential, and oscillating around the CP-conserving value. The most simple cosmological model assumes that the Universe at this early epoch is dominated by radiation. This is the minimal assumption because this needs to be the case during Big Bang Nucleosynthesis, which follows shortly after. However, other options are possible and plausible, like a period of matter (moduli) domination, kination (i.e., an epoch dominated by the kinetic energy of a field), or even an additional phase of inflation.

\begin{figure}[htbp]
\begin{center}
\includegraphics[width=\textwidth]{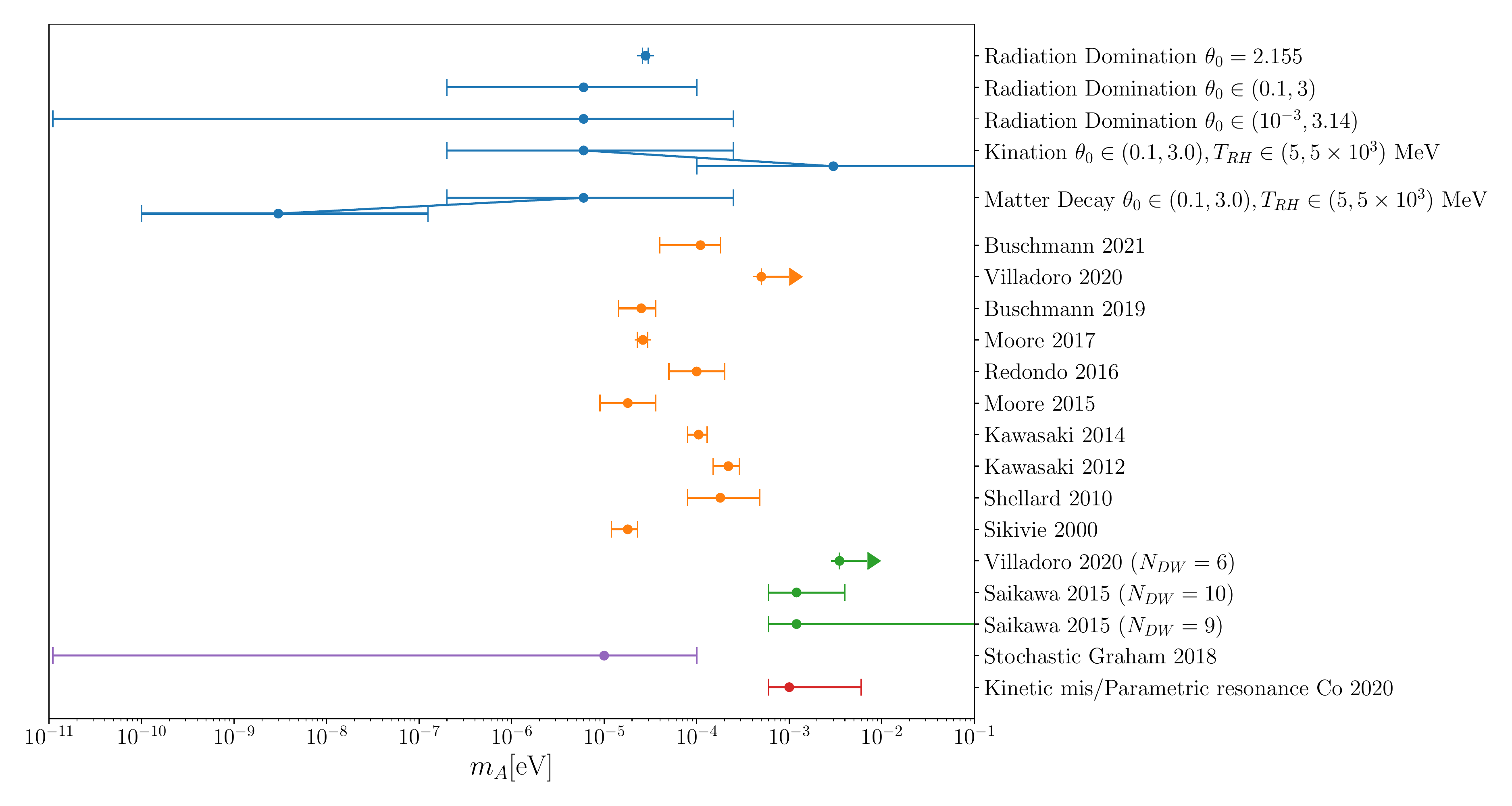}
\caption{The axion DM mass in different scenarios: preinflation (blue)~\cite{Borsanyi:2016ksw,Visinelli:2018wza}, post-inflation $N_{\rm DW}=1$ (orange)~\cite{Buschmann:2021sdq,Gorghetto:2020qws,Buschmann:2019icd,Klaer:2017ond,Ballesteros:2016xej,Fleury:2015aca,Kawasaki:2014sqa,Hiramatsu:2012gg, Wantz:2009it,Sikivie:1999sy}, and $N_{\rm DW}>1$ (green) \cite{Gorghetto:2020qws,Ringwald:2015dsf}, Stochastic (purple) \cite{Graham:2018jyp} and Kinetic misalignment (red) \cite{Co:2019jts}.}
\label{compi}
\end{center}
\end{figure}

Figure \ref{compi} shows a compilation of the range of predictions for the axion DM mass in different cosmological scenarios. From top to bottom, we first list the prediction for the axion DM mass in the pre-inflation scenario if the initial condition (initial angle $\theta_i$) is known and fixed to 2.155, and the Universe is radiation dominated. The prediction has an uncertainty of a few percent \cite{Borsanyi:2016ksw}. However, if we recognize our uncertainty in the initial angle in $(0,\pi)$ the uncertainty in the axion DM mass becomes huge. Avoiding the ``tuned regions" of $0$ and $\pi$ by a factor of 0.1 spans the range $m_a\in(2\times 10^{-7},10^{-4})$ eV and by 0.01 the range booms to  $m_a\in(<10^{-11},3\times 10^{-4})$ eV. Furthermore, if axions become DM during a period of kination, these ranges shift up in mass by more than 2 orders of magnitude, whereas if they become DM during a period of matter domination they are similarly shifted to lower axion masses. The upward or downward shift in axion mass depends on how long the non-standard period of expansion is before entering radiation domination again.

Next, the post-inflationary predictions for a domain wall number of 1 ($N_{\rm DW}=1$) in the radiation domination case are shown.
At the moment, they span the $(10^{-5},10^{-3})$ eV range, due to the computational/theoretical uncertainty that pertains the extrapolation to the required tension of the ensuing cosmic global strings. In Matter decay (or kination) cosmologies, we expect the DM mass to be shifted to lower (or higher) masses by similar amounts to the pre-inflationary case. The case $N_{\rm DW}>1$ is excluded by observations, unless some explicit PQ-symmetry broken is included, and the ensuing parameters give additional freedom to the DM mass. However, the general trend in this scenario is that the expected axion mass is distributed over a broader range than in the $N_{\rm DW}=1$ case. 
Finally, we report two scenarios discovered recently. In the stochastic scenario, axion DM is due to quantum fluctuations in a small-scale inflation. In the kinetic misalignment one, the initial conditions of the axion field are prepared with huge kinetic energy by a PQ-breaking sector at very early times. The expansion rate during inflation and the new fields allow much freedom for the axion DM mass in these cases.  

Recent years have seen an increased interest to design experiments to cover the post-inflationary scenario, and collaborations like ADMX and CAPP have drafted their experimental campaigns as a slow increase in mass from the sweet-spot for cavity haloscopes, $\sim $\textmu eV towards the $10^{-4}$ eV range. Many new experiments such as MADMAX \cite{MADMAX:2019pub}, ALPHA \cite{ALPHA:2022rxj}, ORGAN \cite{McAllister:2017lkb}, CADEx \cite{Aja:2022csb}, or BREAD \cite{BREAD:2021tpx} aim directly at higher masses, motivated in the context of post-inflationary predictions, with a standard radiation dominated early Universe.

Extending the cavity haloscope concept to lower masses by $\sim$ one order of magnitude below ADMX is also strongly motivated. On the one hand, the range of axion DM predictions reaches down to these values. For instance, the pre-inflationary scenario with an initial angle of $\theta_i\sim 0.1$, hardly a fine tuned value, predicts axion DM in this range. Even in the post-inflationary scenario, a small period of matter domination can easily shift the axion DM mass prediction below the \textmu eV scale without clashing with any cosmological observation. Any cosmological event diluting axion DM or creating new entropy will have much of the same effect and will similarly shift the constraints. 

On the other hand, we have the experimental ups. First, the scanning rate of a cavity haloscope scales as 
%\be 
$\frac{d\log m_a}{dt} \propto m_a^{-3} V^2 T_{\rm sys}^{-2}$.
%\ee
Therefore, lower masses have much faster scanning rates
owing to not only the $m_a^{-3}$ factor but to larger volumes, and the system noise can also be sensibly smaller at lower frequencies. Second,   
although it might be very difficult to get our hands on decimeter sized multi-T magnets, there are many $O(\textrm{m})$-sized multi-T magnets available that make the sub-\textmu eV axion DM mass range search feasible. 
As an example, the FLASH experiment has been proposed for using the FINUDA magnet (1.200 m$^3$ volume, 1.1 T B-field) to do an axion search around the \textmu eV with KSVZ sensitivity \cite{Flash}.  
Most notably, the next generation of axion helioscopes aims at building a 20 m-long, 5 m-diameter toroidal magnet for the International Axion Observatory (IAXO)~\cite{IAXO:2019mpb}. This magnet, or part of it could be available in the future for axion DM experiments. 
While funds are raised and the community is prepared for IAXO, an intermediate step helioscope, BabyIAXO, will be build~\cite{IAXO:2020wwp}. It will feature 2 bores 0.7 m wide and 10 m-long, with 2 T B-field for a critical test of IAXO technology and the solar axion search. This magnet can be available to the axion community for axion DM searches.

In this paper, we discuss challenges and opportunities in using the BabyIAXO magnet to host axion DM experiments.
The paper is structured as follows:
In Section \ref{sec:cryo}, we review the BabyIAXO project and propose a concept for a dedicated cryostat that could be inserted into one of the warm bores of BabyIAXO.
Section \ref{sec:cavity_des} describes the design of the RADES cavities, including a possible tuning mechanism. 
The envisaged readout and DAQ of the system are described in Section \ref{sec:rf}, and sensitivity prospects in the axion parameter space computed in Section \ref{sec:sensitivity}.
Section \ref{sec:BASE-DM} sketches an additional concept to search for axion CDM at lower masses, with lumped element LC circuits.
Finally, we conclude in Section \ref{sec:conc} and show the overall reach of the proposed concepts.

\section{BabyIAXO and a proposal for a cryostat inside its warm bore 
\label{sec:cryo}
}

\subsection{BabyIAXO}
The BabyIAXO helioscope is the smaller demonstrator of the larger IAXO helioscope and serves as a test version of the components at a technically representative scale  \cite{IAXO:2020wwp}.
Simultaneously, it will work as a fully functional helioscope with sensitivities beyond CAST, the option to detect axions or Axion Like Particles (ALPs) and a rich physics case \cite{IAXO:2019mpb,Dafni:2021mqa,DiLuzio:2021qct}.
BabyIAXO is currently under construction at DESY in Hamburg.
In addition to the operation as an axion helioscope, BabyIAXO offers excellent properties to be used as a haloscope because it allows for more magnetic volume than any existing setup.
For the helioscope experiment, BabyIAXO will move continuously following the sun.
During the operation as a haloscope in a later stage, BabyIAXO would remain stationary in its parking position.
A vacuum valve close to the helioscope detectors allows to leave the x-ray detectors installed.
The haloscope system can then be inserted into one of the bores of BabyIAXO from the other side.
Beyond the operation of the BabyIAXO magnet the haloscope setups will otherwise be independent experiments.
This enables a second detection strategy for the IAXO collaboration with a different target parameter space and small additional costs due to the shared magnet and infrastructure of BabyIAXO.
To allow this, the cryostat needed for the haloscope experiments should be independent and insertable into the bore, while allowing for maximal resonant cavity volumes.
Additionally, an independent cryostat will allow comprehensive testing of the cryostat and haloscope experiments prior to their insertion in the magnet thus maximizing time for physics.
There is no liquid helium cryogenic infrastructure available at the currently planned BabyIAXO site, thus a dry cryostat was considered.
The layout of the specially designed cryostat fulfilling the stringent requirements is sketched in the following and in~\cite{Siodlaczek2022}.

\subsection{Cryostat design}
\begin{figure}[ht]
\centering
\includegraphics[width=\textwidth]{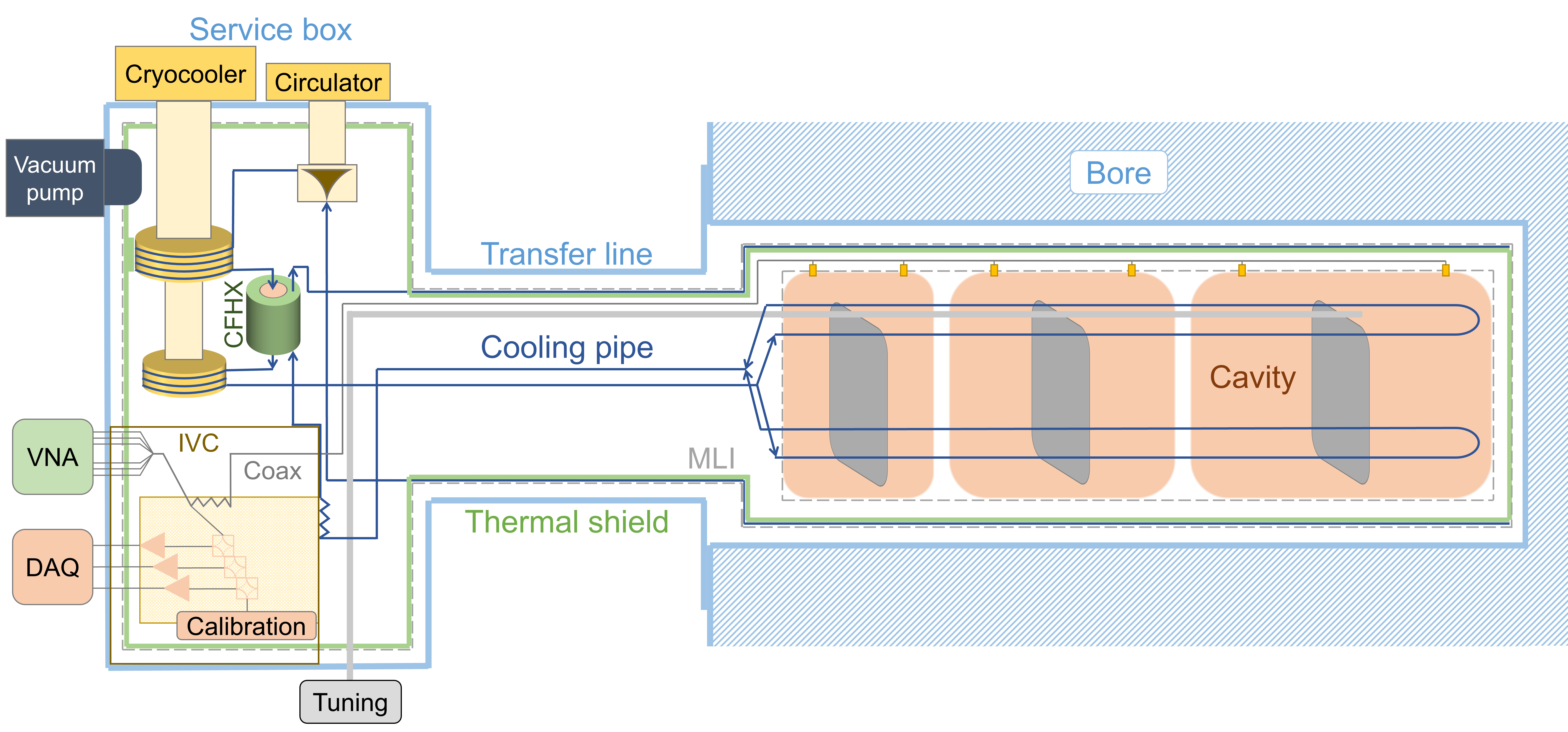}
\caption{Sketch of the preliminary design (not to scale) for the haloscope cryostat at BabyIAXO including the RF system components, cooling system components, and the way of connecting those components with the cooling pipes.
Three resonant cavities are shown as example. The coupling of the cryostat to the individual resonators can be done similarly for the different haloscope designs. 
Picture from~\cite{Siodlaczek2022}.}
\label{CryostatHaloscope}
\end{figure}

The cryostat design is based on cooling via cryocoolers and a closed helium circulation loop, which was considered an effective solution without cryogenic infrastructure.
The layout of the cryostat is sketched in Figure \ref{CryostatHaloscope}.
 In this section we take as example of resonators a set of cavities equipped with mechanical tuning rods. 
The resonant cavities inside the BabyIAXO bore are surrounded by a thermal shield and sustained by a support structure made out of G10 epoxy-fiberglass.
Ball transfer units on the structure enable the insertion of the support structure, the thermal shield, and the resonators into the bore.
The cavities and the thermal shield are cooled by small cooling pipes that are brazed onto the surface and in which supercritical helium is circulated.

A 30 layer Multi-Layer Insulation (MLI) blanket covers the thermal shield and a 10 layer MLI blanket protects the cavities.
Each cavity is connected to two coaxial cables at two ports and to the tuning and coupling mechanisms, see section \ref{sec:cavity_des}.
These mechanical subsystems consist of stepper motors at room temperature, piezoelectric actuators at low temperature, and hollow rods for transmitting the movement.
The helium transports the heat to the cryocoolers in a service box outside the external magnetic field and a cryogenic circulator regulates the helium mass flow rate.
A \SI{2}{\m} long transfer line extends the vacuum chamber of the bore to the service box and the cooling pipes, the coaxial cables, and the tuning rods are routed through the transfer line.
The service box hosts all the auxiliary equipment of the cooling system as well as part of the radiofrequency (RF) equipment such as the low noise amplifiers (LNA), switches, and calibration systems.

An effective counter-flow heat exchanger (CFHX) is required next to the cryocoolers in series and the cryogenic circulator to enable the operation of this cooling loop.
The CERN Cryolab is doing extensive research on effective mesh-based counter-flow heat exchangers, which 
we plan to use in the present cryostat in view of their encouraging performance~\cite{Onufrena2021,Onufrena2021a,Onufrena:2022nwx}.
The RF equipment at low temperature is placed into an instrumentation vacuum chamber and cooled by thermal contacts to a copper plate onto which the cooling pipe is brazed in a meander.
This additional instrumentation vacuum chamber (IVC) acts as a second vacuum environment and facilitates the maintenance of the RF equipment to which the data acquisition system (DAQ) and vector network analyzer (VNA) are connected. \\

The heat load in the present cryostat is estimated via numerical modeling to assess the feasibility of the cooling system and the resulting temperature levels.
This modeling also enables design optimization to minimize the resulting heat load.
More information on the estimation of the heat load can be found in~\cite{Siodlaczek2022}.
The total heat load is dominated by thermal radiation due to the huge surface areas inside the bore.
For the estimation of this, the MLI model from Riddone~\cite{Riddone1997} is applied and multiplied with an application factor, which considers the MLI performance for complex surface geometries with multiple degradation points. This approach provides an indicative value for the heat load and determines the relation between heat load and steady-state cavity temperature.  
Additionally, the heat load through the coaxial cables is estimated with a 2D model, taking into account the heat transfer between the layers of the cable.
The solid heat conduction through the G10 support structure in the bore and the walls of the instrumentation vacuum chamber is calculated and the resulting heat load at the different locations is considered.
The heat produced by the RF equipment is assessed according to the respective specification sheets and the assumed operating scheme.
Lastly, the heat load of the tuning system is dominated by the hollow tuning rods due to thermal conduction and thermal radiation along their important length and is numerically estimated with a 1D model.
The estimation of the total heat loads for a system with a thermal shield at \SI{50}{\K}  and bore at \SI{4.5}{\K} corresponds to \SI{37.3}{\watt} at \SI{50}{\K} and \SI{1.69}{\watt} at \SI{4.5}{\K}.
Of the total \SI{37.3}{\watt}, \SI{30.2}{\watt} occur in the bore and \SI{7.1}{\watt} in the service box at \SI{50}{\K}.
Analogously, \SI{1.38}{\watt} and \SI{0.31}{\watt} at \SI{4.5}{\K} can be assigned to the heat load in the bore and the service box, respectively.
The origin of the heat loads at the service box and bore are presented in Table~\ref{tab:HeatLoadSummary}.
\begin{table}[ht]
\centering
\caption{Summary of heat loads in steady state at the two temperature levels of the cryostat.
The heat load values, the sources and the locations of their occurrence are specified.
}
\begin{tabular}{l S[table-format=2.1]S[table-format=2.1]S[table-format=2.1] S[table-format=1.2]S[table-format=1.2]S[table-format=1.2]}
\hline
{Source}    &\multicolumn{3}{c}{Heat load @\SI{50}{\K} (\si{\watt})} &\multicolumn{3}{c}{Heat load @\SI{4.5}{\K} (\si{\watt})} \\
			&{Service box}  &{Bore} &{Total}    &{Service box}  &{Bore} &{Total} \\
\hline
{Radiation / MLI}		    &3.3 &29.3 &32.6 	&{-}   &1.31 &1.31  \\
{Coaxial cables}	        &2.4 &{-}  &2.4 	&0.18  &{-}  &0.18  \\
{Support structure \& IVC}	&1.0 &0.9  &1.9 	&0.06  &0.05 &0.11  \\
{Electrical equipment}	    &0.1 &{-}  &0.1 	&0.07* &{-}  &0.07* \\ 
{Tuning system}			    &0.3 &{-}  &0.3 	&{-}   &0.02 &0.02  \\
\hline
{Total}					    &7.1 &30.2 &37.3 	&0.31  &1.38 &1.69  \\
\hline
\end{tabular}

\smallskip
\parbox[t]{\textwidth}{\footnotesize
	* Partly dynamic and not steady state heat load.}
\label{tab:HeatLoadSummary}
\end{table}

The proposed cooling system has been described in a thermodynamic model which enables an investigation and optimization of the system.
The complex dependences among the effectiveness of the cryocooler, the effectiveness of the cryogenic circulator and the heat load mean that a numerical study is essential.
Thus, the cryocooler type, the number of cryocoolers, the appropriate cryogenic circulator, and the sequence of the components in the cooling system was studied and their influence on the cavity temperature was evaluated.
The most promising mass flow rate and pressure regime were chosen based on the model.
These optimizations concluded in a cooling system consisting of two Cryomech\textsuperscript{\textregistered} PT420 cryocoolers~\cite{CryomechPT420} connected in series, one Stirling Cryogenics\textsuperscript{\textregistered} Noordenwind cryogenic circulator~\cite{StirlingCryo}, one counter-flow heat exchanger specifically produced by the CERN Cryolab, cooling pipes with an inner diameter of \SI{4}{\mm}, and a helium flow rate of \SI{0.5}{\g\per\s} at \SI{20}{\bar} absolute pressure.
The optimized sequence of the components is shown in Figure \ref{CryostatHaloscope}.
According to these calculations, a cavity temperature of about \SI{4.6}{\K} and an RF equipment temperature of about \SI{4.8}{\K} can be reached.
More information on the design of the haloscope cryostat at BabyIAXO can be found in~\cite{Siodlaczek2022}.

\section{Cavity design}
\label{sec:cavity_des}
\begin{figure}
    \centering
    \includegraphics[scale=0.7]{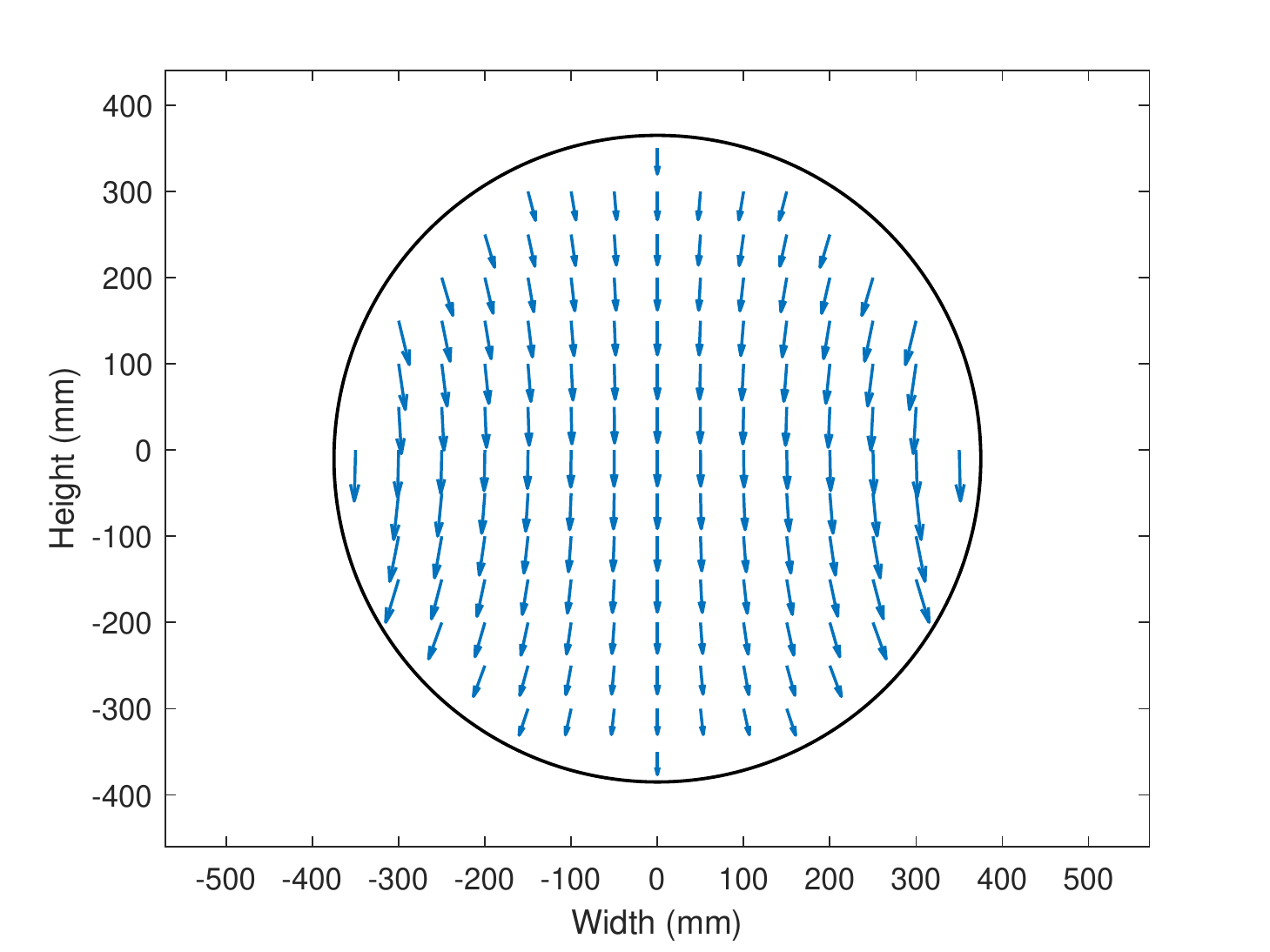}
    \caption{Static magnetic field distribution in a BabyIAXO magnet section in the $xy$ plane, $z$ being the bore axis. This determines the selection of the cavity mode yielding a maximum form factor.}
    \label{fig:magnet_field}
\end{figure}

In this section, the design of the resonant haloscope and ancillary systems such as tuning or coaxial-cavity coupling mechanisms are outlined. These designs are highly constrained  by the space available in the bore of the magnet and the external magnetic field pattern, which depends on the magnet type (solenoid or dipole) and other characteristics of the magnet. BabyIAXO uses a racetrack magnet with a cross-section magnetic field distribution depicted in Figure \ref{fig:magnet_field}.
We assume that, after the installation of the cryostat inside the magnet bore, the available volume corresponds to a cylinder of 60~cm in diameter and 10~m in length.

\subsection{Operational parameters}
As in any resonant haloscope design, our goal is optimizing the sensitivity of the experiment when scanning a large frequency interval by means of a tuning procedure.

The expression for the detected power needed to evaluate the sensitivity of the experiment is \cite{kim:2020,Jeong:2022akg}

\be
  P_d = \kappa g_{a \gamma} ^2 \frac{\rho_{\rm DM}}{m_a} B_e^2 C V \frac{Q_lQ_a}{Q_l+Q_a} ~.
\ee
or, simplifying for the usual case when $Q_a \gg Q_l$,
\be
\label{eq:Pg}
    P_d = \kappa g_{a \gamma} ^2 \frac{\rho_{\rm DM}}{m_a} B_e^2 C V Q_l ~.
\ee

The key haloscope parameters for achieving the aforementioned goal are the volume of the cavity ($V$), the form factor of the chosen resonant mode ($C$), the loaded quality factor for that mode ($Q_l$) and the coupling factor ($\kappa$), or altogether, the product $\kappa Q_l C V$. The form factor is given by \cite{Jeong:2022akg}
\be \label{eq:form_factor}
    C = \frac{|\int \vec{B_e} \cdot \vec{E}\: \mathrm{d}V|^2}{\int |\vec{B_e} |^2 \mathrm{d}V  \int \varepsilon_r |\vec{E}|^2\: \mathrm{d}V} ~,
\ee \\
where $\Vec{B_e}$ is the static magnetic field, $\Vec{E}$ the dynamic electric field induced by the axion-photon conversion, and $\varepsilon_r$ the relative electrical permittivity in the haloscope.
The remainder of the parameters in Equation \ref{eq:Pg} are the unknown axion-photon coupling $g_{a \gamma}$, the dark matter density $\rho_{\rm DM}$, the axion mass $m_a$, the quality factor of the axion resonance $Q_a$, and the static magnetic field $B_e$, which are independent of the haloscope design. \\

Both $\kappa$ and $Q_l$ can be written in terms of the coupling coefficient $\beta$ as
\be \label{eq:kappa}
    \kappa = \frac{\beta}{1+\beta} ~,
\ee
\be \label{eq:Ql}
    Q_l = \frac{Q_0}{1+\beta} ~,
\ee
where $Q_0$ is the cavity unloaded quality factor for the chosen resonant mode. In this way we separate the parameters depending exclusively on the cavity geometry and the material used for its construction ($V$, $C$ and $Q_0$) from those depending on the geometry and position of the coaxial-cavity coupling ($\beta$).\\

Therefore, the maximum detected RF power ($P_d$) for one frequency is obtained by maximizing the product $\frac{\beta}{(1+\beta)^2} Q_0 C V$. The first factor is maximum for $\beta = 1$, the so-named critical coupling regime, where the maximum energy of the cavity (half of the total energy) is extracted. Assuming that we can design a coaxial-cavity coupling working in the critical coupling regime, the remaining product to be optimized is $Q_0 C V$, which, as already stated, depends only on the cavity geometry and the chosen resonant mode. We name this product as the single–frequency performance figure of the unloaded cavity  ($\Pi_c = Q_0 C V$), and, by taking into account the coaxial cable–cavity coupling, we can define a performance figure for the loaded cavity:
\be \label{eq:FOM2}
    \Pi_{cc} = \frac{\beta}{(1+\beta)^2} \Pi_c
\ee

which, under critical coupling conditions, becomes

\be
    \Pi_{cc} = \frac{\Pi_c}{4}
\ee

In this proposal we assume this single frequency power optimization, that is, critical coupling ($\beta=1$), but we remark that in the final prototype, the optimal $\beta$ will be used in order to maximize the scanning rate $\( \frac{d m_a}{dt} \propto \frac{\beta^2}{\left( 1 + \beta\right)^3} Q_0 C^2 V^2 \)$, required to reach a desired signal-to-noise to set an upper limit on $g_{a\gamma}$ \cite{kim:2020,Jeong:2022akg}. In a first approach, $\beta=2$ is the optimal value for this second criterion, but taking into account \cite{kim:2020} and the high $Q_0$ expected at these low frequencies $\( \frac{Q_a}{Q_0} \sim 10 \)$, a higher value of $\beta$ may be considered that may substantially reduce the data-taking time obtained here.\\

We now present in Section \ref{sec:single_cav} a single cavity design that fills the entire volume available in one bore, to provide an insight on the parameters at play.
We then propose in \ref{sec:multiple_cav} and \ref{sec:multiple2_cav} realistic multiple cavity designs.

\subsection{Single cavity design} \label{sec:single_cav}
\begin{figure}[h]
    \centering
    \includegraphics[scale=0.2]{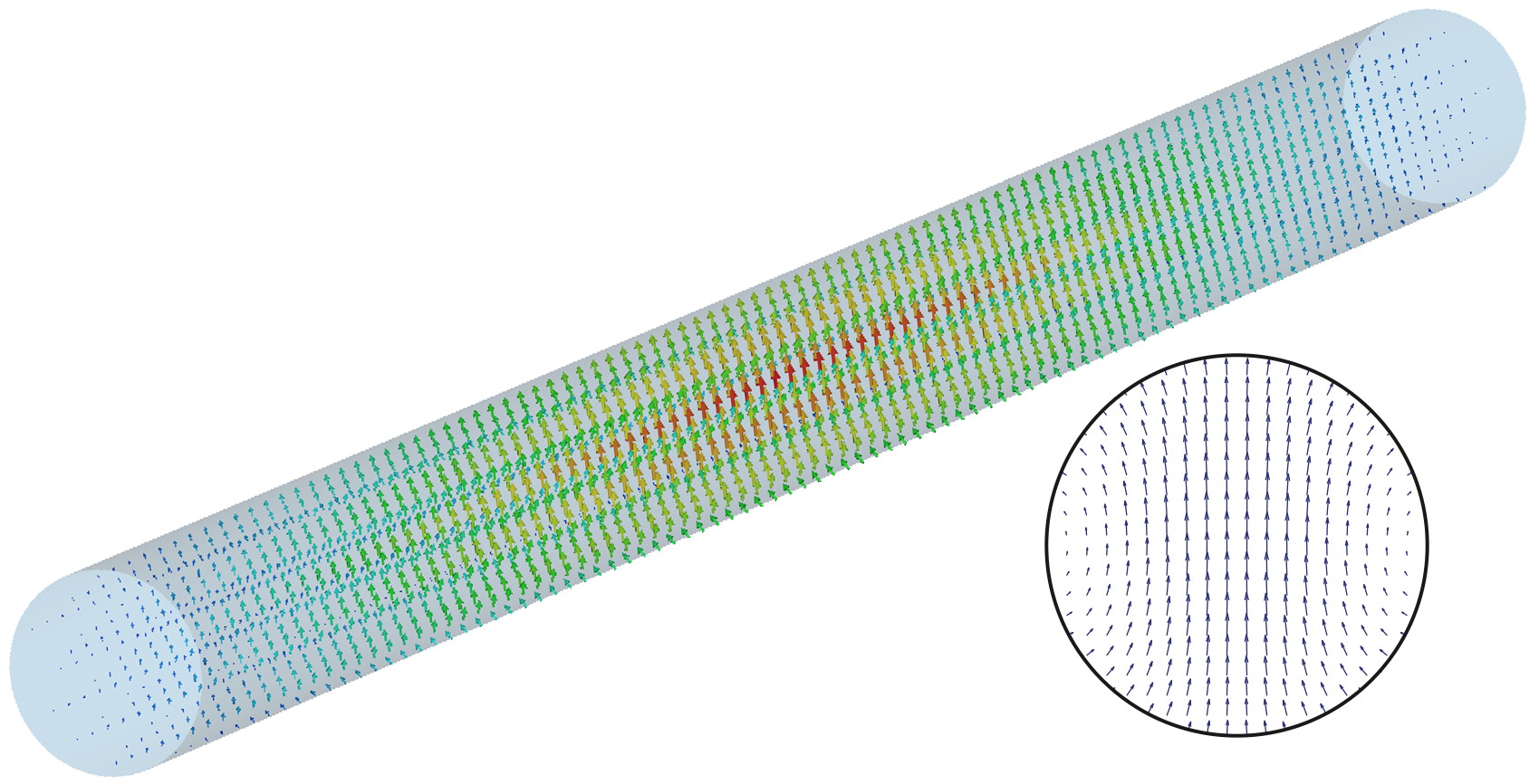}
    \caption{Electric field pattern for the $TE_{111}$ mode in a cylindrical cavity and cross-section detail.}
    \label{fig:TE111_E}
\end{figure}

\begin{figure}[h]
\centering
\begin{subfigure}[b]{0.49\textwidth}
         \centering
         \includegraphics[width=0.99\textwidth]{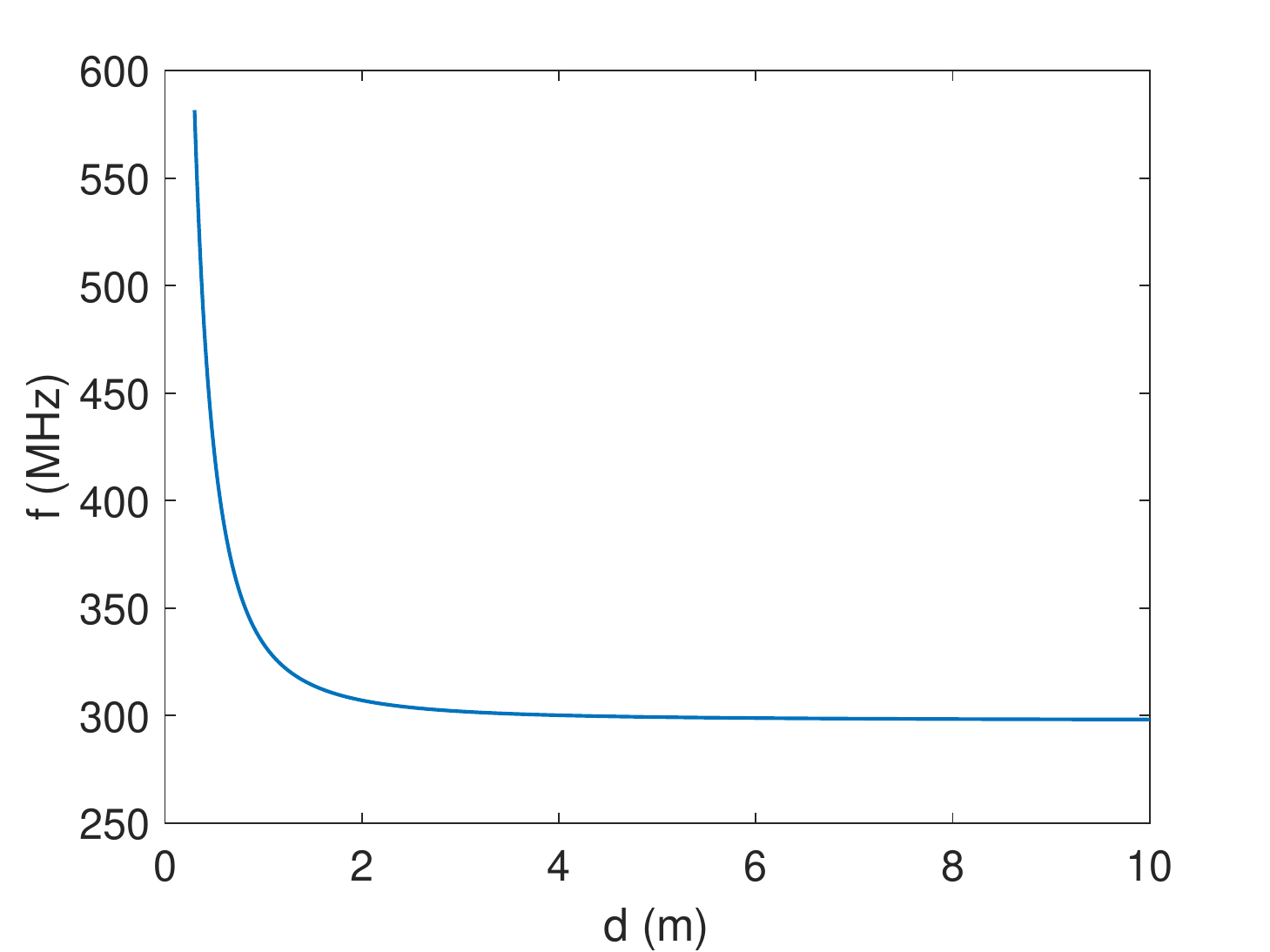}
         \caption{}
          \label{fig:fr_TE111}
\end{subfigure}
\hfill
\begin{subfigure}[b]{0.49\textwidth}
         \centering
         \includegraphics[width=0.99\textwidth]{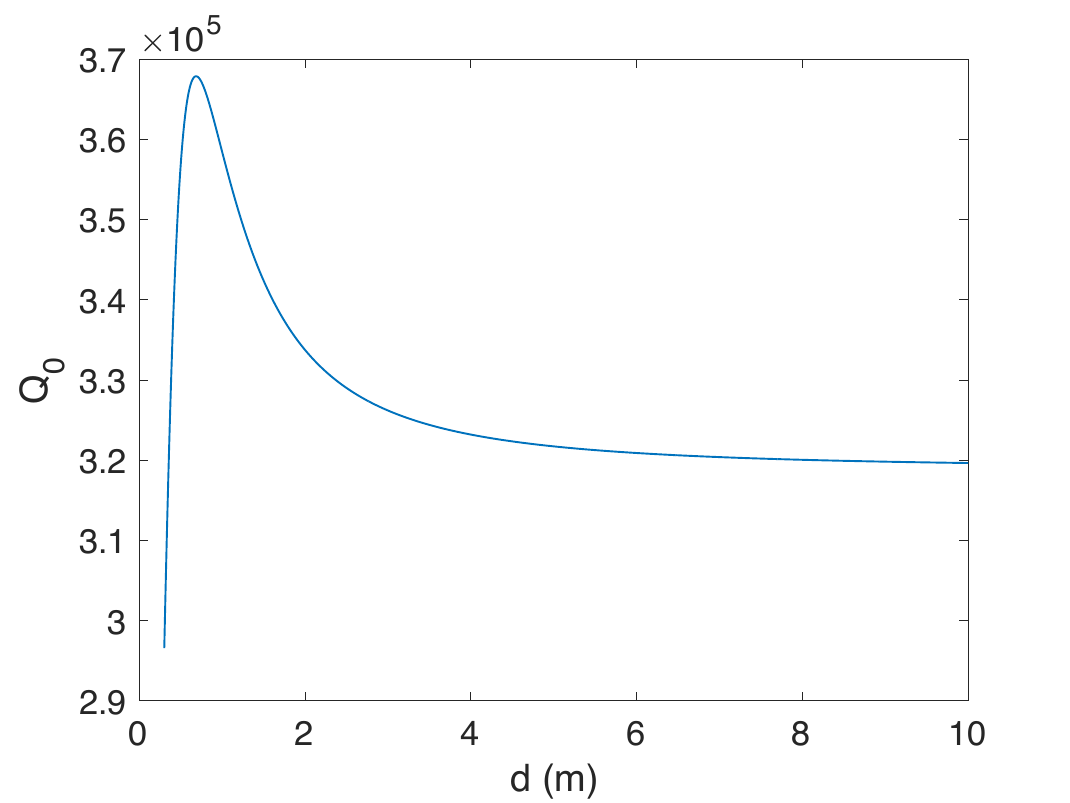}
         \caption{}
         \label{fig:Q0_TE111}
\end{subfigure}
\caption{(a) Resonant frequency, and (b) unloaded quality factor for $TE_{111}$ mode. Both plots correspond to a copper cylindrical cavity with a $59$~cm inner diameter as a function of length $d$.}
\end{figure}

The straightforward design for a resonant cavity which maximizes the available volume  within the cryostat is a cylindrical–shaped single cavity with a diameter of 60~cm and a length of 10~meters, with an inner diameter of 59~cm (assuming a wall thickness of $0.5$~cm) and total volume of 2,734 liters. Taking into account the spatial pattern of the external magnetic field shown in Figure \ref{fig:magnet_field}, the proper resonant mode for detecting the axion is the $TE_{111}$ mode, whose electric field is depicted in Figure \ref{fig:TE111_E} for the whole cavity.
The distribution in a transversal section is also included for the sake of clarity. The form factor for this mode is $C=0.658$, assuming the external magnetic field of BabyIAXO magnet. The resonant frequency can be calculated from \cite{Collins2001}
\be\label{fr_cyl}
    f_{TE_{nlq}} = \frac{c}{2\pi\sqrt{\varepsilon_r}}\sqrt{\(\frac{p^{'}_{nl}}{a}\)^2+\(\frac{q\pi}{d}\)^2} ~,
\ee
where $n$, $l$ and $q$ are the number of variations of the mode in azimuth, radial and axial variables, respectively, $c$ is the speed of light in vacuum, $\varepsilon_r$ is the relative electrical permittivity within the cavity (in this case 1), $p^{'}_{nl}$ is the $l$-th root of the derivative of the Bessel function of order $n$, $a$ is the cavity radius, and $d$ is its length. For the $TE_{111}$ mode this resonant frequency is
\be\label{TE111_fr_cyl}
    f_{TE_{111}} = \frac{c}{2\pi}\sqrt{\(\frac{1.8411}{a}\)^2+\(\frac{\pi}{d}\)^2} ~,
\ee
which is plotted in Figure \ref{fig:fr_TE111} for an inner diameter of $59$~cm and variable length. For the maximum cavity length $d=10$~m, the resonant frequency is located at $298.17$~MHz, which corresponds to an axion mass of $m_a=1.234$~\textmu eV.

The unloaded quality factor for $TE_{nlq}$ modes due to conductor losses are given by \cite{Collins2001}  
\be\label{Q0_TEnlq}
    Q_{0TE_{nlq}} = c\sqrt{\frac{\pi\mu_0\sigma}{f}}\frac{\(1-\(\frac{n}{p^{'}_{nl}}\)^2\)\(p^{'2}_{nl}+\(\frac{q\pi a}{d}\)^2\)^{\frac{3}{2}}}{2\pi\(p^{'2}_{nl}+\frac{2a}{d}\(\frac{q\pi a}{d}\)^2+\(1-\frac{2a}{d}\)\(\frac{nq\pi a}{p^{'}_{nl}d}\)^2\)} ~,
\ee
where $\mu_0$ is the magnetic permeability of the vacuum and $\sigma$ is the electrical conductivity of the cavity walls. For $TE_{111}$ mode and a diameter of $59$~cm, the behavior of $Q_0$ with the cavity length is shown in Figure \ref{fig:Q0_TE111}. $Q_{0TE_{111}} = 3.2\times10^5$ for the maximum length $d=10 \, $m .

\begin{figure}[h]
\centering
\begin{subfigure}[b]{0.49\textwidth}
         \centering
         \includegraphics[width=0.99\textwidth]{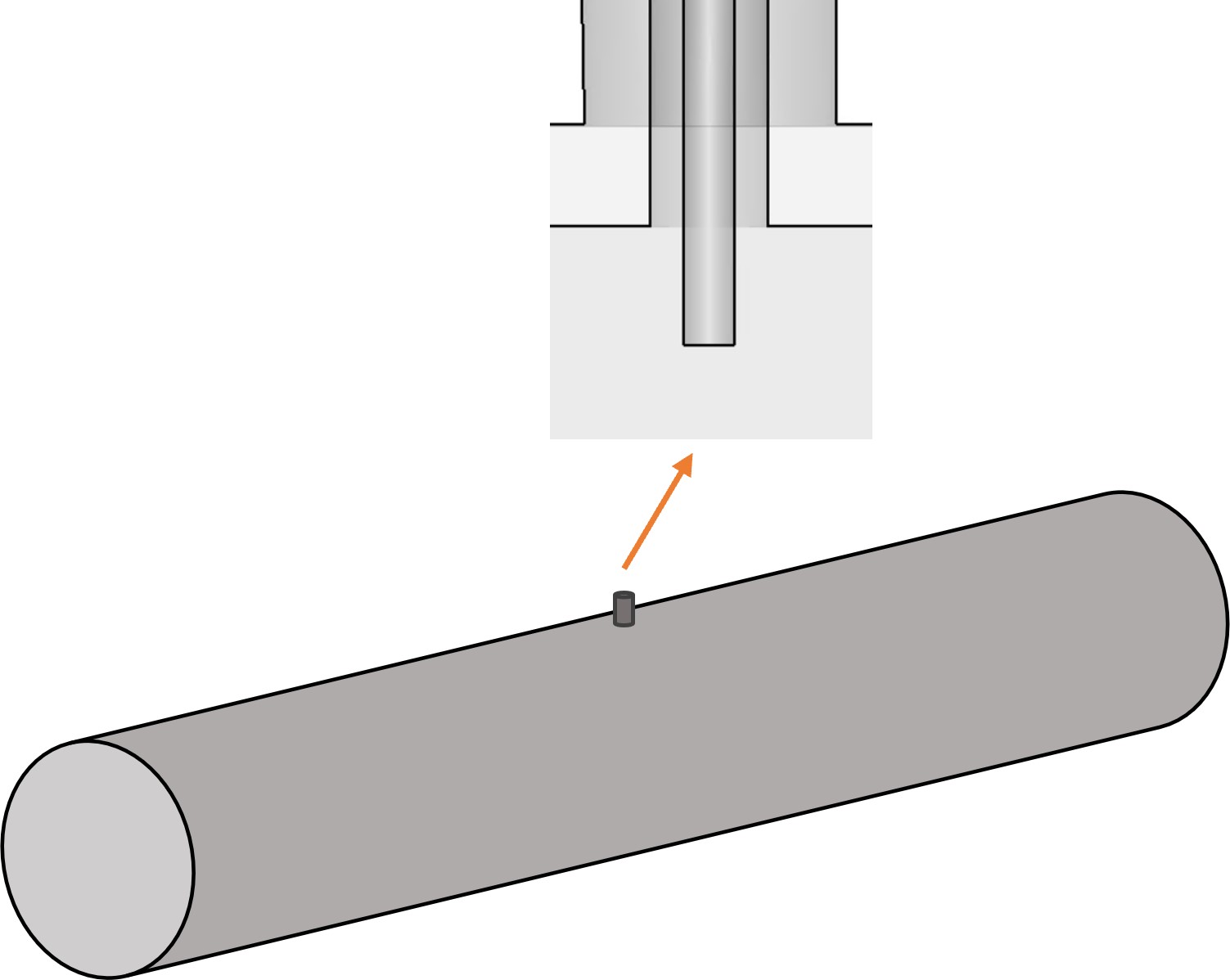}
         \caption{}
          \label{fig:elec_coupling}
\end{subfigure}
\hfill
\begin{subfigure}[b]{0.49\textwidth}
         \centering
         \includegraphics[width=0.99\textwidth]{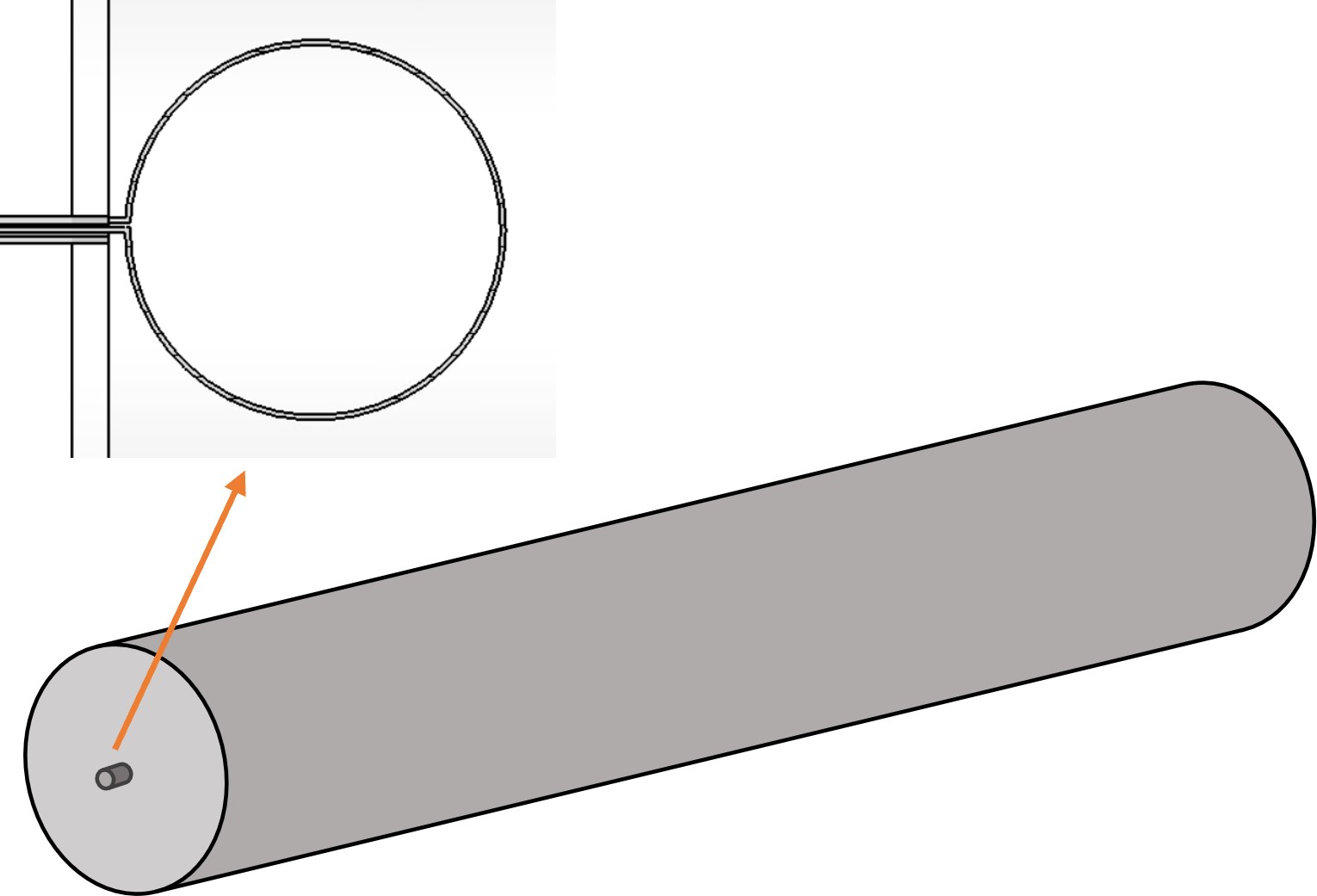}
         \caption{}
         \label{fig:mag_coupling}
\end{subfigure}
\caption{Mechanisms for coupling of the $TE_{111}$ mode in a cylindrical cavity: (a) monopole probe for electric coupling, and (b) loop probe for magnetic coupling.}
\end{figure}

A final important aspect in the single cavity design is the type of coupling (electric or magnetic) used for extracting energy from the cavity. Electrical coupling is normally implemented by means of a straight monopole coaxial probe in a zone where the electric field is parallel to the inner conductor and maximum (see Figure \ref{fig:elec_coupling}). For the $TE_{111}$ mode this maximum occurs at the point $z=d/2$ (see Figure \ref{fig:TE111_E}).
\\

Magnetic couplings employ a loop probe located on a zone with maximum magnetic field crossing the loop. This occurs for $TE_{111}$ mode in the centre of any of the cavity caps, as depicted in Figure \ref{fig:mag_coupling}. Clearly, magnetic coupling is the preferred solution in our case, since it saves five meters of coaxial cable length and, therefore, the attenuation associated with that distance. Moreover, it reduces the clearance that would be necessary on the side of the cavity to accommodate the electric coupling mechanism.\\

However, this single-cavity design has a main drawback: the high clustering effect with modes varying in the $z$ direction will hinder the mass exploration, generating a great number of mode crossings during the tuning process. Moreover, other practical reasons related with the large size and weight of a 10 m cavity advise to use shorter cavities. 
\\
\subsection{Multiple cavities} \label{sec:multiple_cav}
\begin{figure}[ht]
\centering
\includegraphics[width=0.85\textwidth]{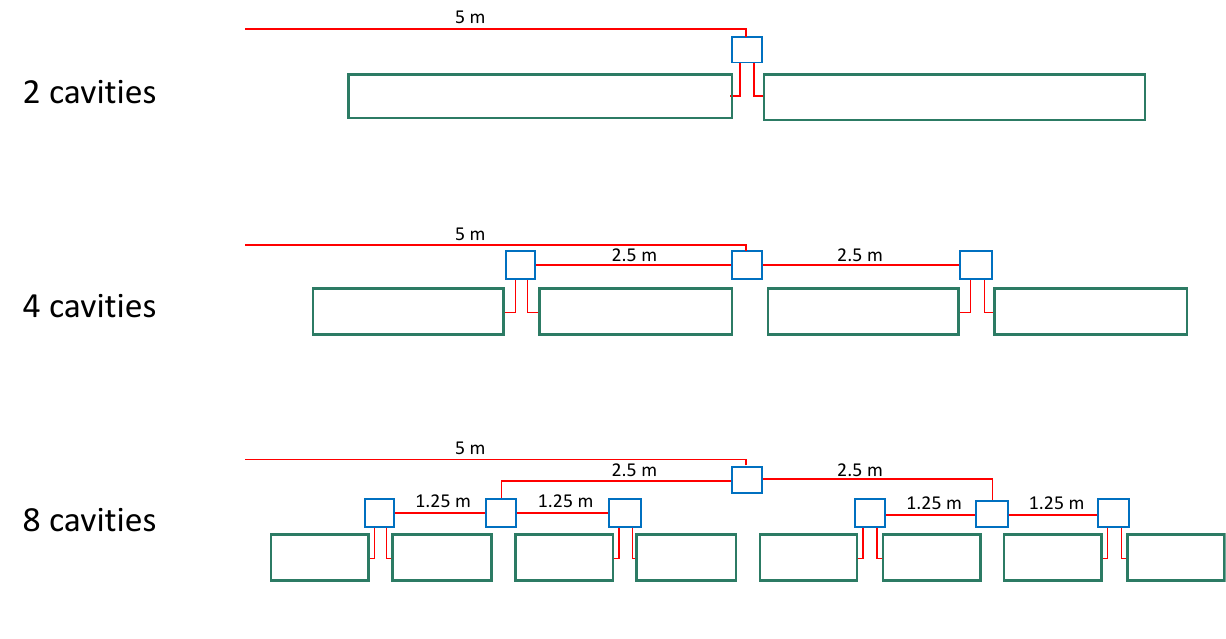}
\caption{Set-ups for coherent sum of signals from $N$ cavities with magnetic coupling, where $N$ is a power of 2. Green color for the cavities, blue for the sum device (combiner) and red for coaxial lines.}
\label{fig:comb_mag_coup}
\end{figure}

In order to reduce these disadvantages, the alternative to the single long cavity is the coherent sum of $N$ shorter cavities (see Figure \ref{fig:comb_mag_coup}), each one with an extracted power $P_{d_i}$ and with a total detected RF power given by
\be\label{total_Pd_eq}
    P_d = N \, P_{d_i} ~.
\ee
This is the maximum power, neglecting internal small losses, a microwave power combiner can obtain, following the principle of energy conservation. If the input signals are not in phase, the output power will decrease.
A more accurate calculation of the total detected power must take into account the attenuation produced in the coaxial cable from the extraction point of each cavity to the final combiner (final sum point). Depending on the type of coupling (electric or magnetic) and the number of employed cavities, the length of the necessary cables and also the attenuation and phase change will vary.\\

\begin{figure}[ht]
\centering
\includegraphics[width=0.85\textwidth]{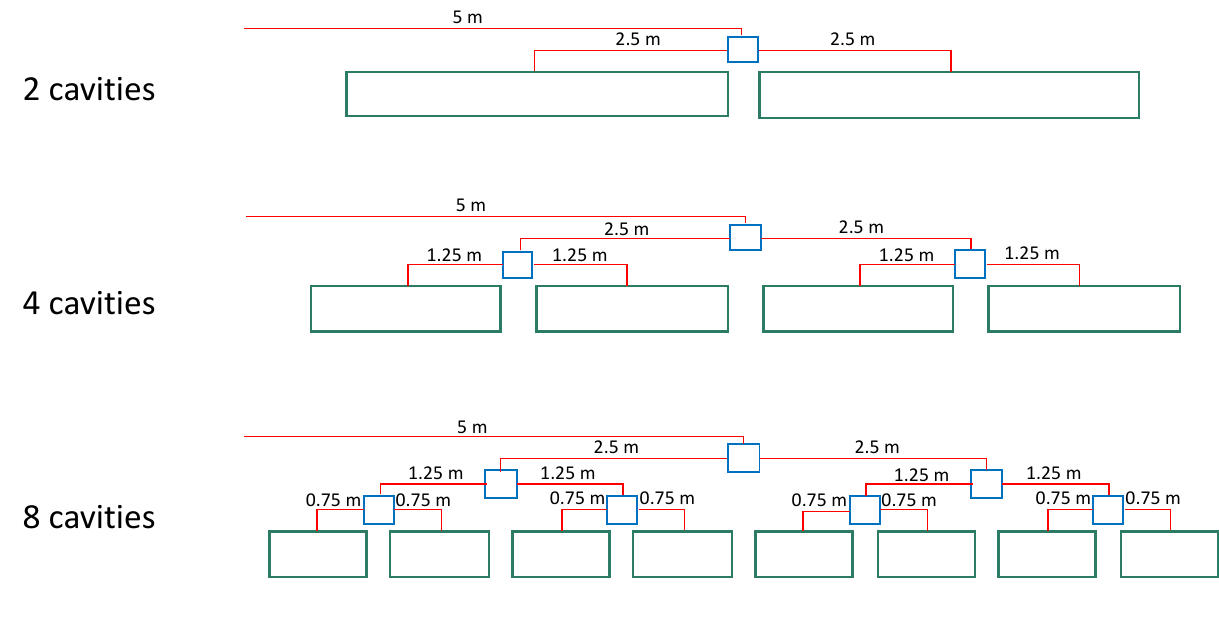}
\caption{Set-ups for coherent sum of signals from $N$ cavities with electrical coupling, where $N$ is a power of 2. Green color for the cavities, blue for the sum device (combiner) and red for coaxial lines.}
\label{fig:comb_elec_coup}
\end{figure}

In this case, since the coherent sum requires the same phase change (that is, the same length of coaxial cables for the signals that are added), a possible set-up for two, four and eight cavities is depicted in Figure \ref{fig:comb_mag_coup}, where the approximate cable length is indicated, assuming magnetic coupling. Similar set-ups can be designed for electrical coupling, which are shown in Figure \ref{fig:comb_elec_coup}. 

\begin{figure}[ht]
\centering
\includegraphics[width=0.6\textwidth]{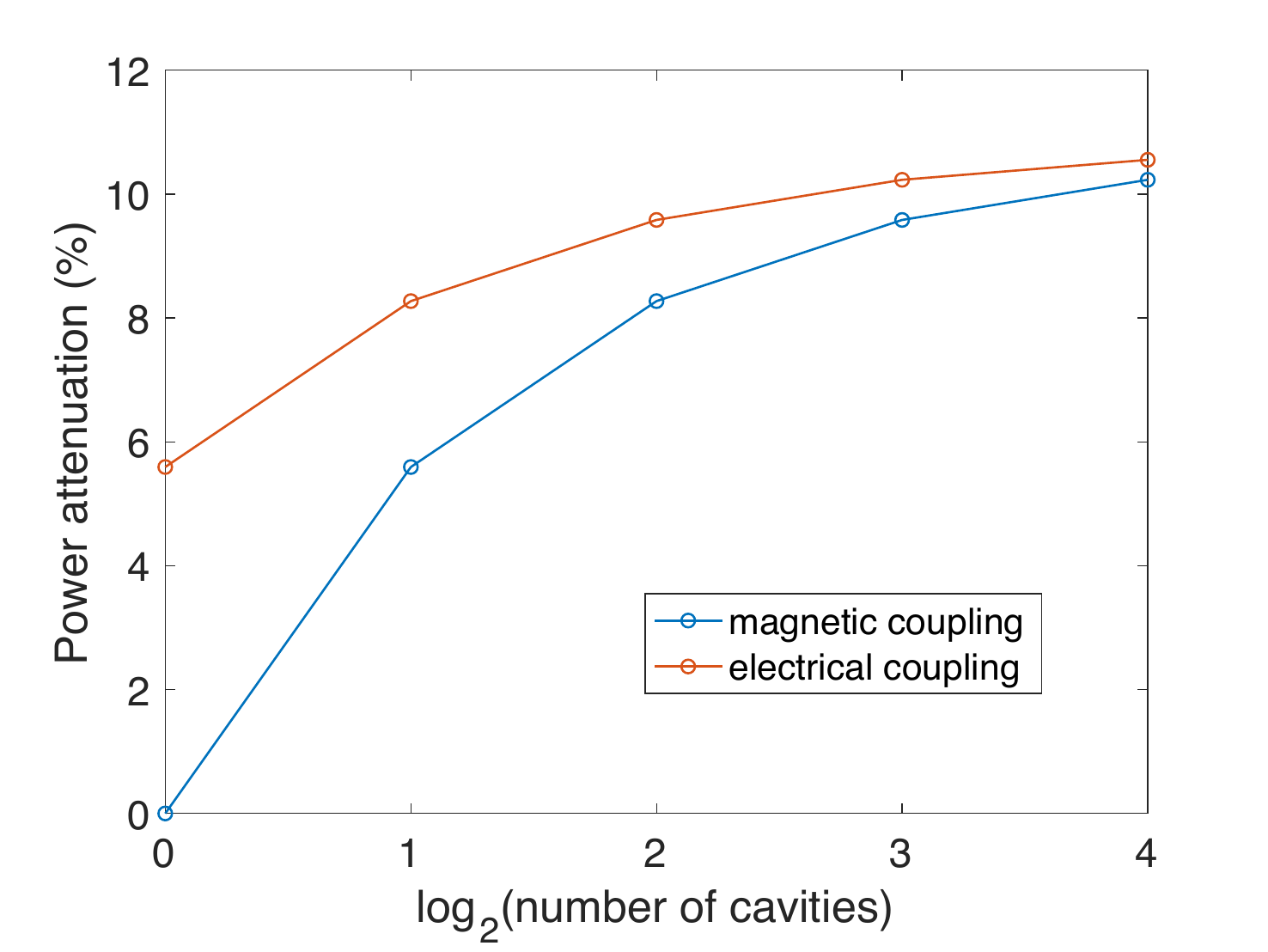}
\caption{Power attenuation ($\%$) depending on the kind of coupling and number of cavities, for a cable attenuation of $0.05$~dB/m. This value is a conservative upper limit calculation for a M17/129-RG401 semi-rigid coaxial cable from Carlisle \cite{Carlisle} working at 4~K.}
\label{fig:power_atten}
\end{figure}

Given the small frequencies in the experiment (in the order of hundreds of megahertz) differences of cable length in the order of 1 cm will produce negligible phase delay among the inputs, and, therefore, negligible effect on the output power. A detailed analysis of coherent sum in haloscopes can be found in \cite{Jeong:2017xqz}.
\\ \\
For the sake of simplicity, we consider negligible in both cases the attenuation of the combiners in comparison with the attenuation of the coaxial lines, although it could be easily considered in the calculations when necessary. Therefore, the total attenuation of the signal up to the bore output for $2^n$ cavities ($n=1,2\dots$) can be calculated in decibels (dB) as a function of the length $d$ of the individual cavities
\be\label{atten_mag_eq}
    A_m(2^n) = \alpha_c\sum^n_{i=1}\frac{d}{2^i}
\ee
for the magnetic coupling case, and
\be\label{atten_elec_eq}
    A_e(2^n) = \alpha_c\sum^{n+1}_{i=1}\frac{d}{2^i}
\ee
for the electrical coupling, where $\alpha_c$ is the coaxial cable attenuation (in dB/m). Figure \ref{fig:power_atten} shows the reduction of detected power in ($\%$) for a cable attenuation of $0.05$~dB/m and for the two different coupling mechanisms.

These results imply that the coupling type and the number of cavities have an impact on the total power detected by the haloscope, and so, on its sensitivity. We therefore define a third single-frequency performance figure that takes this into account:
\be\label{FOM3_eq}
    \Pi_{ccc} = \Gamma\cdot \Pi_{cc} = \Gamma\frac{\beta}{\(1+\beta\)^2}Q_0 \,C \,V ~,
\ee
where $\Gamma=10^{-A_{e/m}/10}$ is the linear gain ($\leq1$ in this case) introduced by the combining system.\\

\begin{figure}[ht]
\centering
\includegraphics[width=0.6\textwidth]{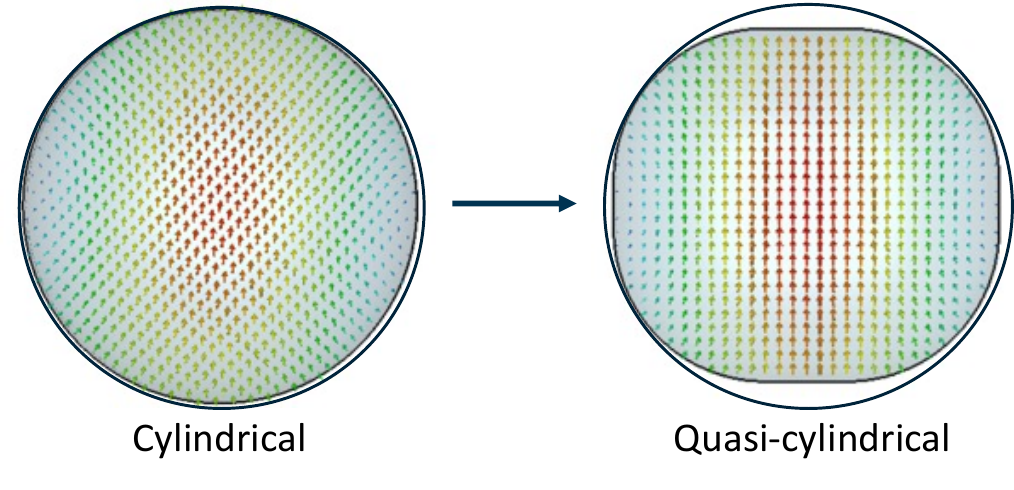}
\caption{Transversal section and electric field ($TE_{111}$ mode) for cylindrical and quasi-cylindrical cavities.}
\label{fig:cyl_and_quasi}
\end{figure}

Another important consequence if multiple cavities are used is the need of additional space for passing the coaxial cables to the second and subsequent cavities. We can reduce the diameter of the cavity or, in order to maximize the volume, slightly modify the shape of the circular cross section in order to liberate some space for cables or even other possible structures (tuning, combiners or structural framing). The concept is illustrated in Figure \ref{fig:cyl_and_quasi}.
This change will slightly modify the electromagnetic field distribution, therefore slightly modifying $f_r$, $Q_0$, and $C$. Table \ref{tab:IndivCav} summarizes the values for these parameters and $\Pi_{ccc}$ assuming critical coupling and the BabyIAXO static magnetic field (see Figure \ref{fig:magnet_field}).\\
 
\begin{table}[ht]
\centering
\caption{Main operational parameters for different configurations of multiple cavities. Magnetic field from BabyIAXO obtained from IAXO collaboration. Assumptions: magnetic and critical coupling, cable attenuation equal to $0.05$~dB/m, lossless combiners, cavities separation of $8$~cm and $0.5$~cm wall thickness.}
\vspace{2 mm}
\begin{tabular}{c c c c c c c c }
\hline
$\#$ cavities & d (m) & $f_r$ (MHz) & V ($\mathrm m^3$) & $Q_0$ & C & $\Gamma$ & $\Pi_{ccc}$ ($ \mathrm m^3$)\\
\hline
1 (cyl) & 9.99 & 298.2 & 2.731 & 3.13e5 & 0.69 & 1 & 1.48e5\\

2 (quasi) & 4.95 & 303.8 & 2.394 & 2.97e5 & 0.63 & 0.944 & 1.06e5\\

4 (quasi) & 2.43 & 308.5 & 2.352 & 3.02e5 & 0.63 & 0.917 & 1.04e5\\

8 (quasi) & 1.17 & 328.3 & 2.252 & 3.19e5 & 0.62 & 0.904 & 1.01e5\\
\hline
\end{tabular}
\label{tab:IndivCav}
\end{table}

From the previous results, a trade-off among mode crossings during tuning, power combination complexity, structural issues, ancillary systems, and $\Pi_{ccc}$ leads to selecting a multiple system with two quasi-cylindrical cavities of $\sim 5$ meters in length, covering frequencies of order 300~MHz.

\subsection{Tuning system design}
\begin{figure}[h]
\centering
\begin{subfigure}[b]{0.57\textwidth}
         \centering
         \includegraphics[width=0.99\textwidth]{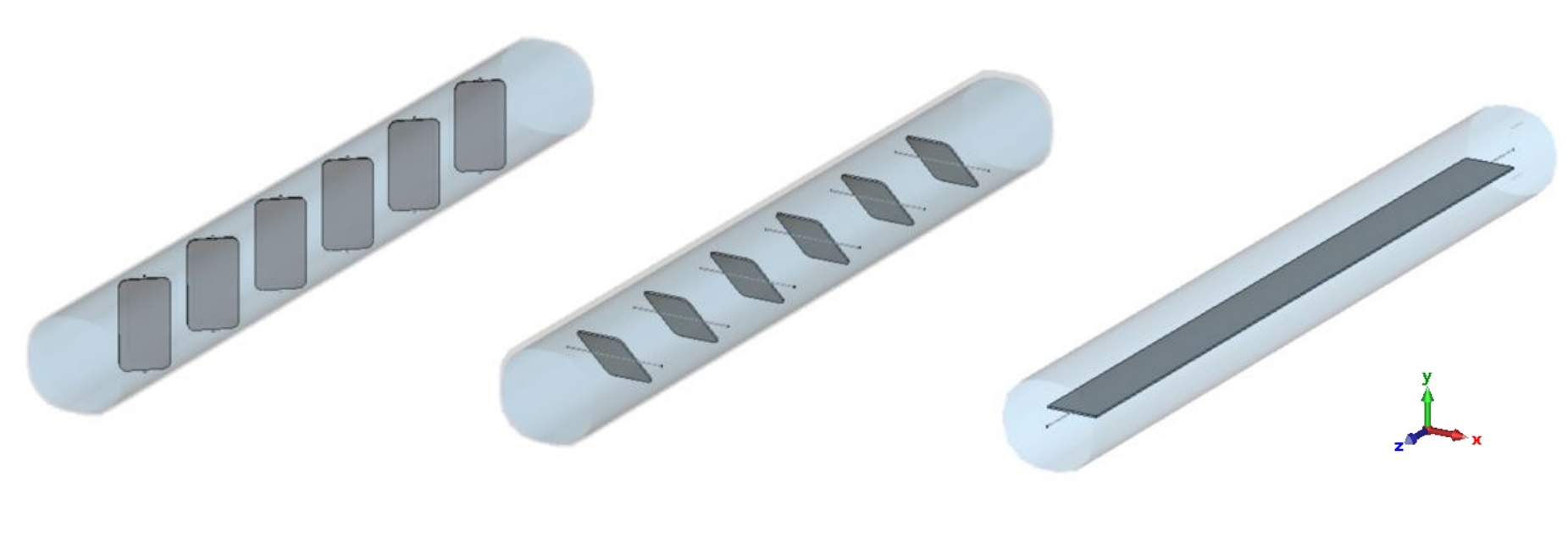}
         \caption{}
          \label{fig:rotating_plates}
\end{subfigure}
\hfill
\begin{subfigure}[b]{0.42\textwidth}
         \centering
         \includegraphics[width=0.99\textwidth]{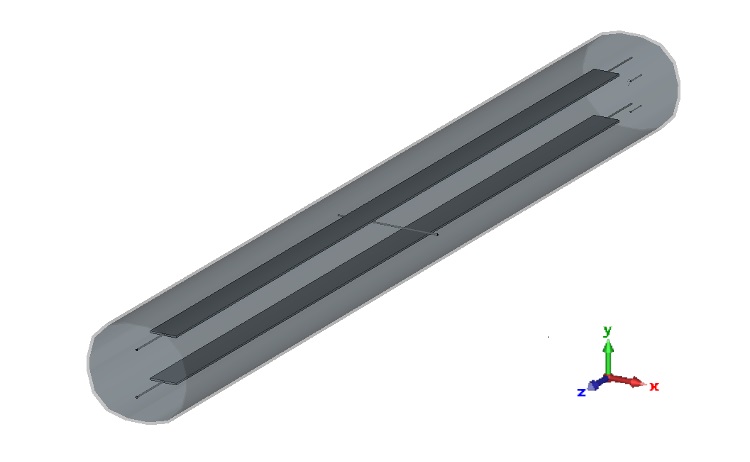}
         \caption{}
         \label{fig:z_axis_2_plates}
\end{subfigure}
\caption{(a) Rotating metallic plate mechanisms in different axes on a 5~m long quasi-cylindrical cavity. From left to right, $x$-axis, $y$-axis and one-plate $z$-axis rotation system for tuning purposes. (b) Two plates $z$-axis rotation on a 5~m long quasi-cylindrical cavity.}
\end{figure}

\begin{figure}[h!]
  \begin{subfigure}[t]{.49\textwidth}
    \centering
    \includegraphics[width=\linewidth]{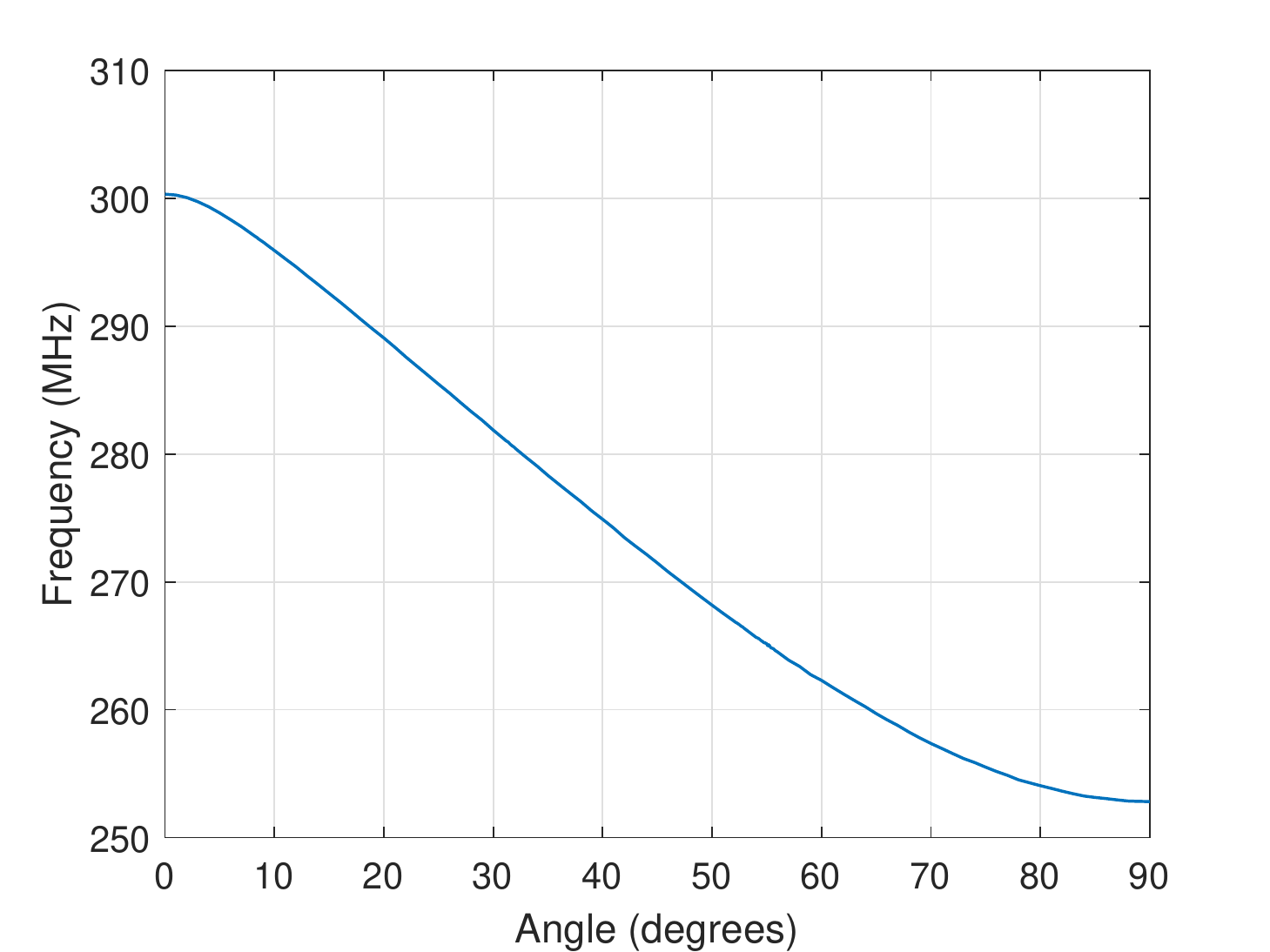}
    \caption{}
    \label{fig:fr_vs_angle}
  \end{subfigure}
  \hfill
  \begin{subfigure}[t]{.49\textwidth}
    \centering
    \includegraphics[width=\linewidth]{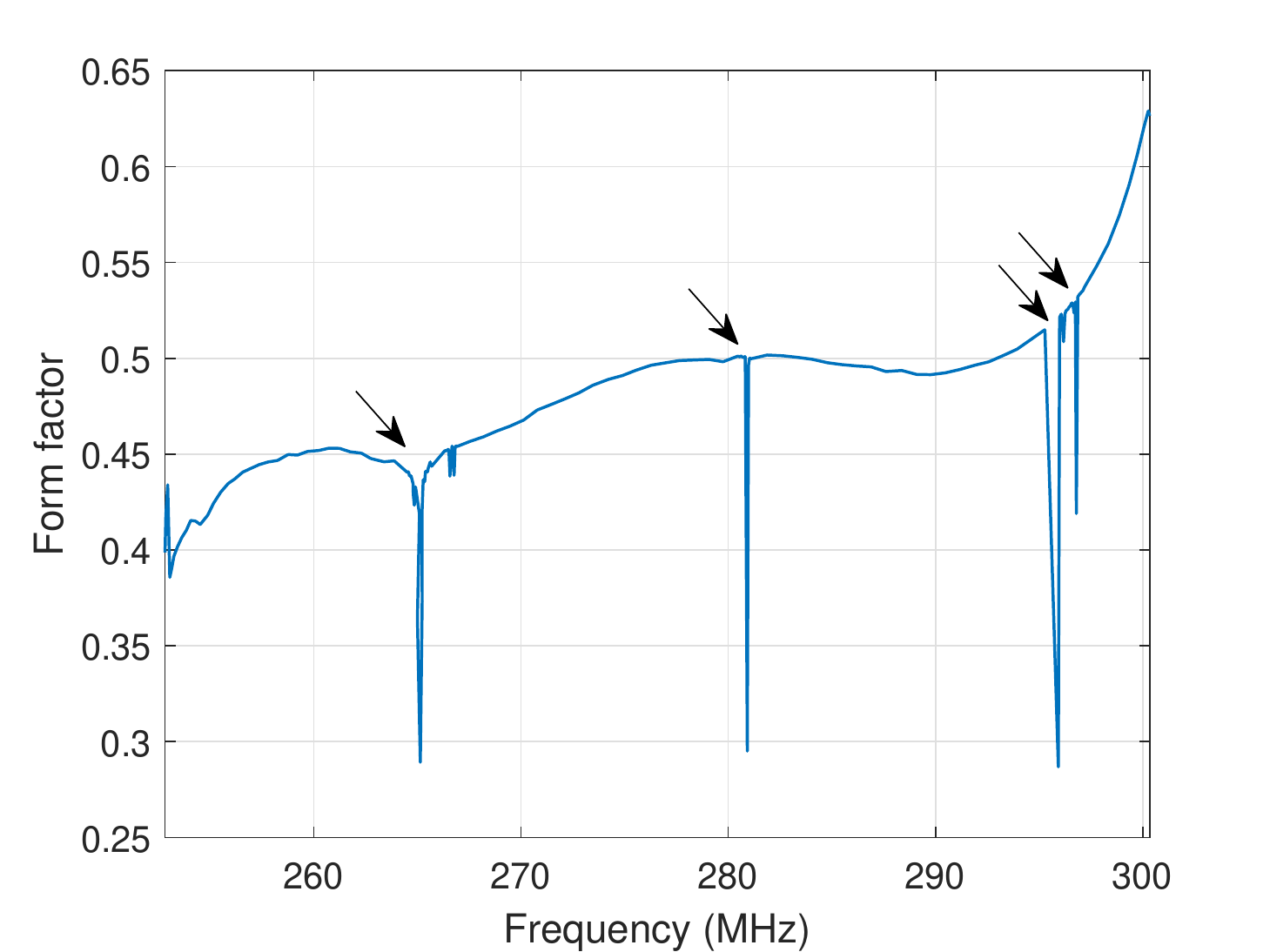}
    \caption{}
    \label{fig:C}
  \end{subfigure}

  \medskip

  \begin{subfigure}[t]{.49\textwidth}
    \centering
    \includegraphics[width=\linewidth]{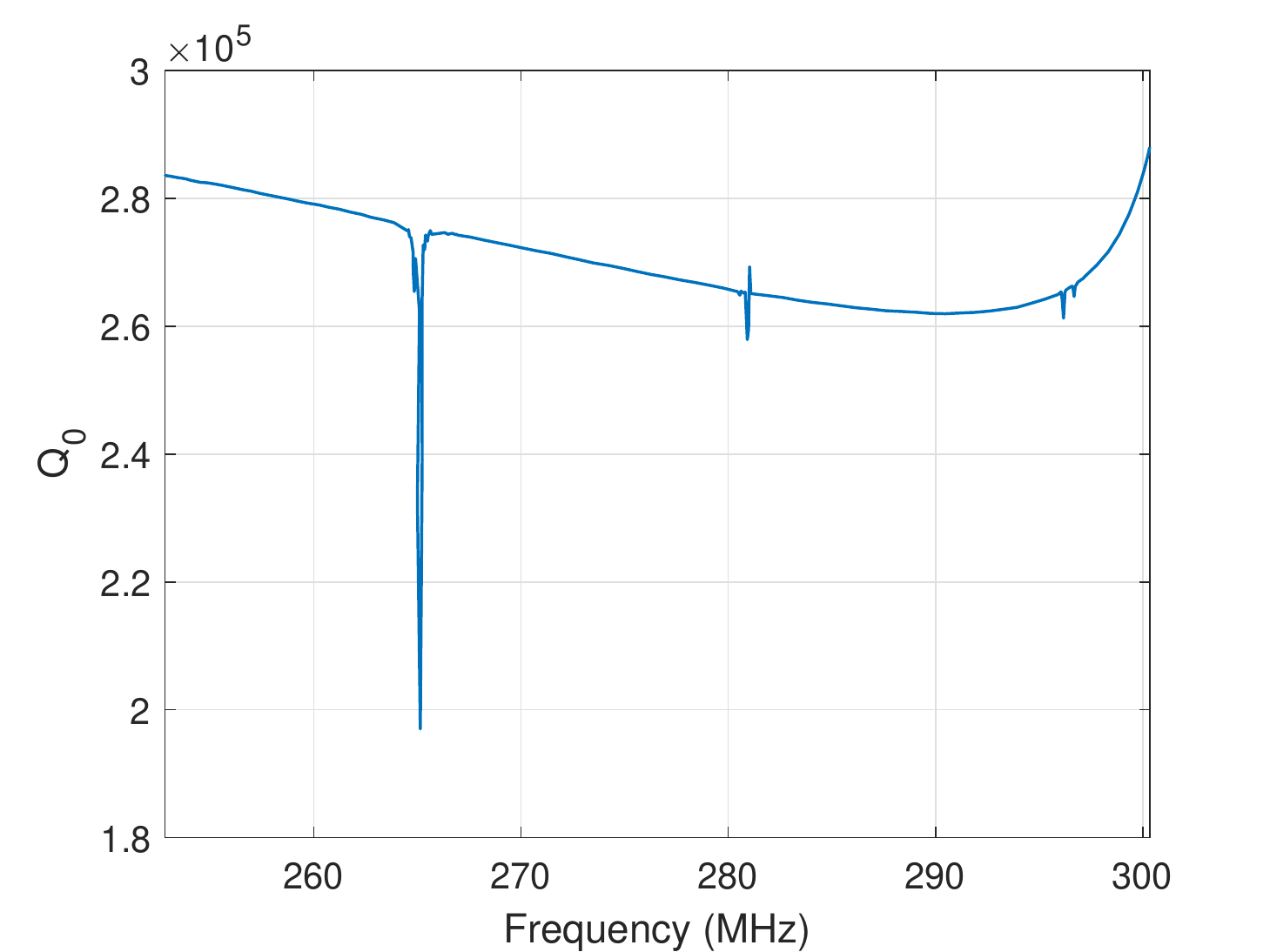}
    \caption{}
    \label{fig:Q0}
  \end{subfigure}
  \hfill
  \begin{subfigure}[t]{.49\textwidth}
    \centering
    \includegraphics[width=\linewidth]{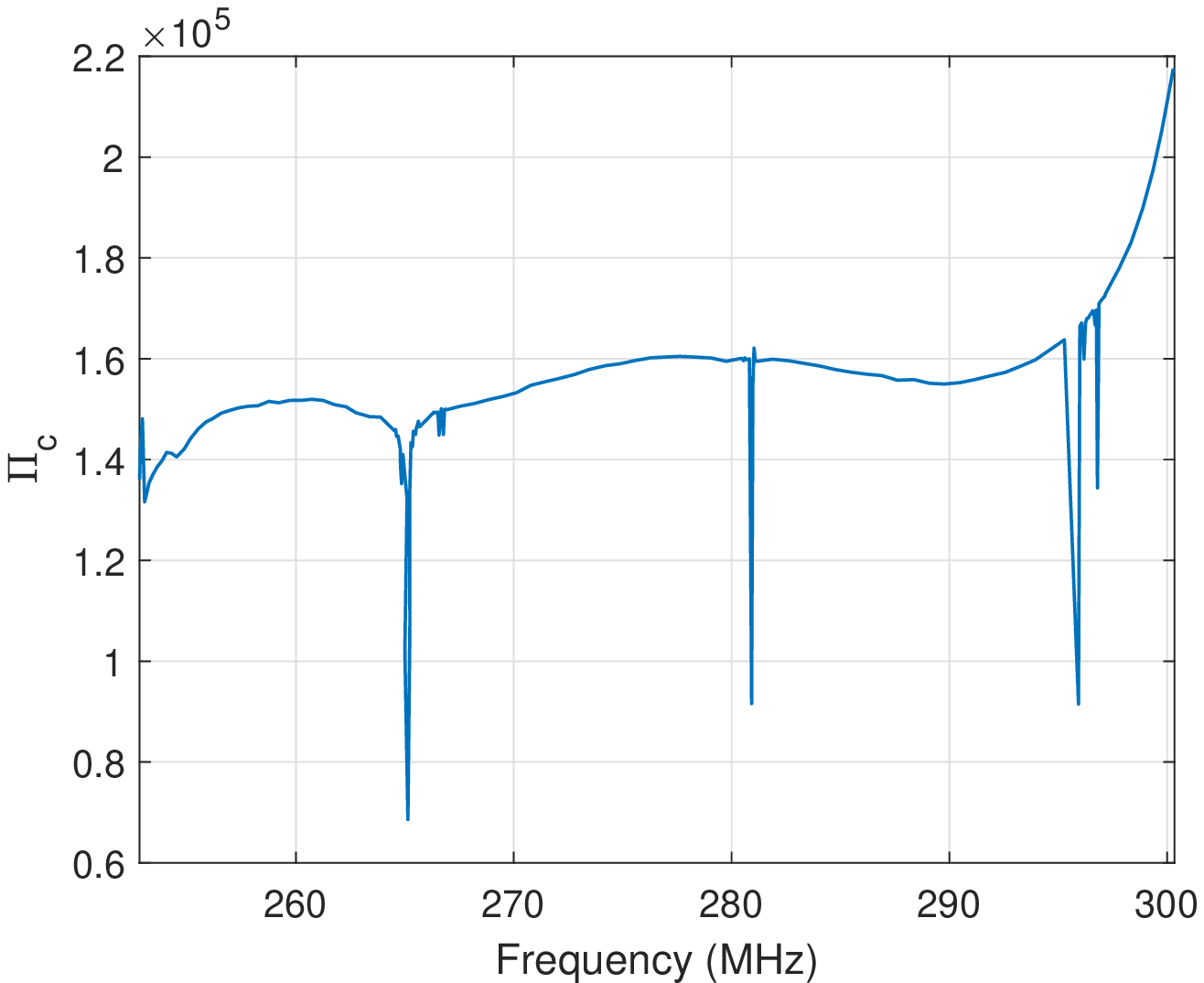}
    \caption{}
    \label{fig:FoM1}
  \end{subfigure}
  \centering
  \caption{(a) Resonant frequency versus angle, (b) Form factor, (c) Unloaded quality factor, and (d) Performance figure $\Pi_c$ versus frequency during the tuning with two plates $z$-axis rotation. Mode crossings are indicated by arrows. Steps of 1 degree are utilized except in the mode crossing zones where steps of 0.1 degrees were used. }
  \label{fig:tuning_params}
\end{figure}

Different mechanical tuning systems have been conceived and simulated, such as a sliding short-circuit along the z-axis, vertical opening of the cavity, dielectric slab insertion, horizontal movement of inner dielectric or metallic plates, and rotary metallic plates. The axial sliding short-circuit is almost insensitive for long cavities, producing a prohibitive reduction of volume if an acceptable frequency range is desired. The vertical opening does not affect the form factor and slightly the volume, providing a wide frequency range, but it requires a very precise mechanism to keep a uniform slot throughout 5 meters, and opens the cavity, with the risk of introducing external interferences. Regarding dielectric slab insertion, not only the leakage (affecting the quality factor) or external interference risk, but the high reduction of the form factor, advise to discard it. The horizontal movement of inner dielectric or metallic plates introduced a high complexity in the mechanical system. A trade-off between simplicity in the tuning mechanism and good performance for frequency range, volume, quality factor and form factor was obtained with rotary metallic plates. This rotation can be performed in the $x$, $y$ or $z$ axis (see Figure \ref{fig:rotating_plates}), but the simplest system (because it does not require parallelization and synchronized movement between different plates) is implemented with a single plate $z$-axis rotation system. Electromagnetic simulations have been performed considering the rotation of one or two plates (see Figure \ref{fig:z_axis_2_plates} for the latter) around the $z$ axis, obtaining much better results in this last configuration. These are presented in Figures \ref{fig:fr_vs_angle} - \ref{fig:FoM1}.

Those mode crossings observed in Figure \ref{fig:tuning_params} must be properly detected during the experiment in order to rule out those blind areas. This will be done by monitoring the $S_{11}$ and $S_{21}$ parameters by means of a VNA connected to the detection port and the weakly coupled port of the haloscope, provided that the interfering mode is conveniently coupled to these ports. Moreover, the use of two rotary plates could allow alternative combinations of plates angles in order to avoid mode crossing in the blind areas produced with the parallel motion of both plates, making possible a complete scanning of the mass range.

\subsection{Variable external coupling system} \label{sec:var_ext_coup}
Since the coupling by means of a loop probe depends on the magnetic flux through the loop, rotating the loop modifies the magnetic coupling. In this way, if the tuning operation produces over-coupled or under-coupled situations, a degree of freedom is available to adjust again the critical external coupling.\\

\begin{figure}[ht]
\centering
\begin{subfigure}[b]{0.49\textwidth}
         \centering
         \includegraphics[width=0.99\textwidth]{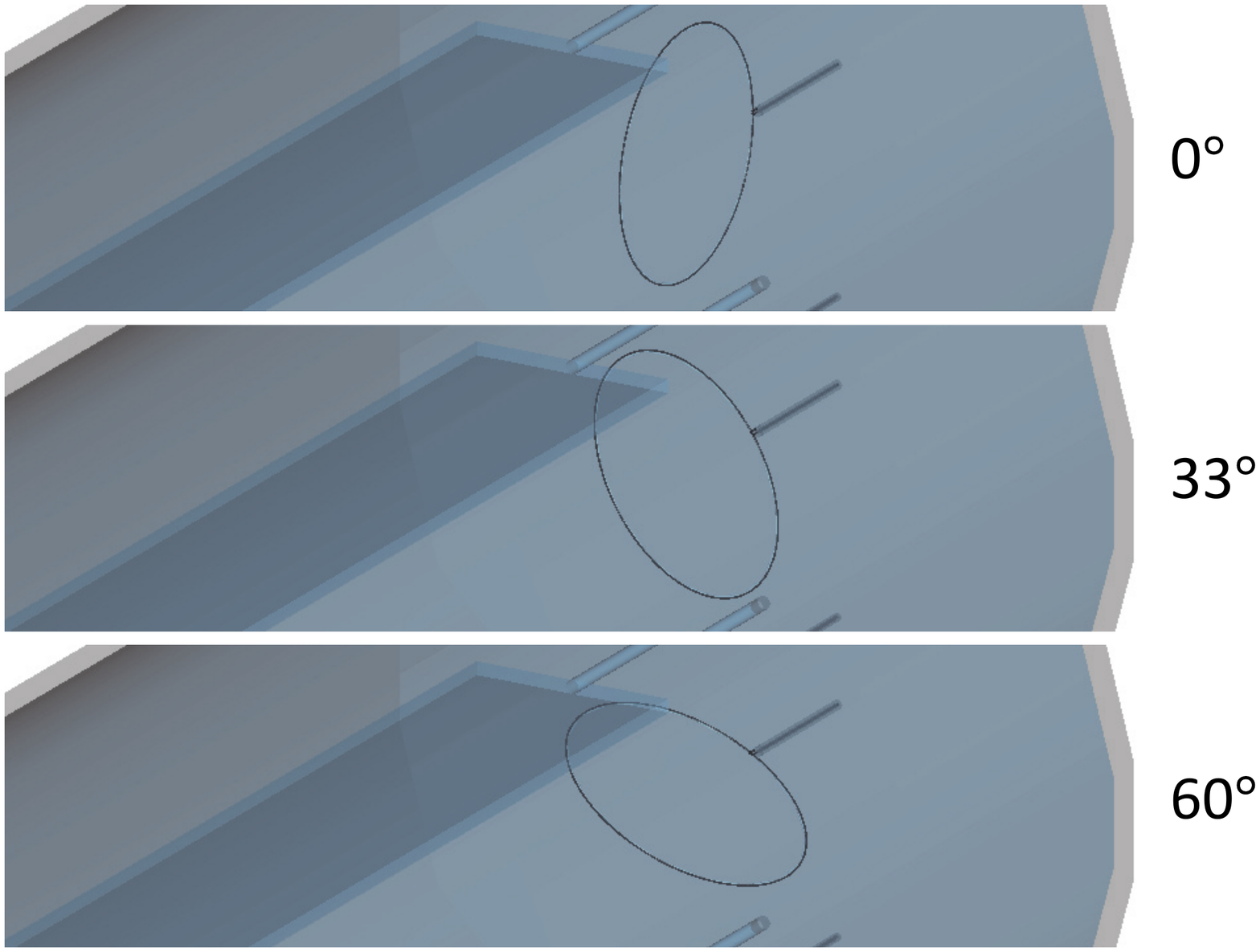}
         \caption{}
          \label{fig:Mechanical_re_coupling}
\end{subfigure}
\hfill
\begin{subfigure}[b]{0.49\textwidth}
         \centering
         \includegraphics[width=0.99\textwidth]{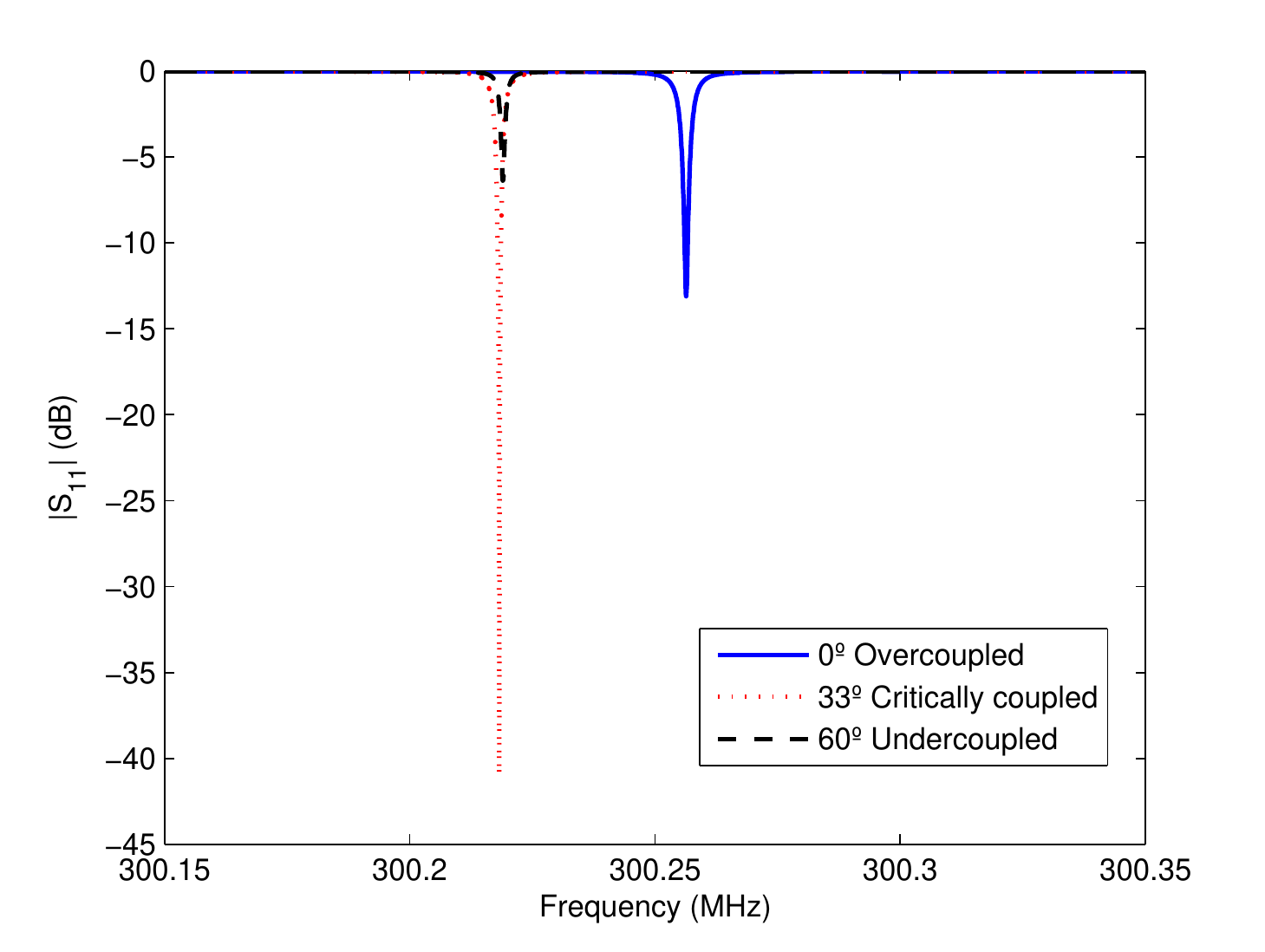}
         \caption{}
         \label{fig:Magnitude_of_the_scattering_parameter_S11}
\end{subfigure}
\caption{(a) Mechanical re-coupling by probe rotation for three different angles (0, 33, and 60 degrees), and (b) magnitude of the scattering parameter $S_{11}$ for different angles of the magnetic probe.}
\end{figure}

Figure \ref{fig:Mechanical_re_coupling} illustrates the probe rotation concept with three different angles ($0$, $33$ and $60$ degrees) of the magnetic probe. The radius of the loop has been fixed to 90~mm for an over-coupled condition when the angle of the loop is 0 degrees (or vertically aligned). As the angle of the loop is varied, Figure \ref{fig:Magnitude_of_the_scattering_parameter_S11} shows that the critically coupled condition is achieved for $33$ degrees. For higher angles, an under-coupled condition is obtained as seen also in Figure \ref{fig:Magnitude_of_the_scattering_parameter_S11}.

\subsection{Alternative concept: exploration of different mass ranges in parallel} \label{sec:multiple2_cav}
In Section \ref{sec:multiple_cav}, the available room in the magnet bore is used for a given mass range exploration, thus optimizing the volume and, therefore, the detected power and the axion-photon coupling sensitivity. An alternative to this concept consists on using two different cavities, each one dedicated to a different mass range. In this way, the volume for each mass region is reduced, but the total explored mass range is increased. An important remark here is that each cavity will need a dedicated heterodyne receiver (cables, LNA and DAQ). Additionally, this set-up requires doubling the measurement time to obtain the same sensitivity and mass range.\\

Taking into account that the single 5 meter long cavity designed in sections \ref{sec:multiple_cav} - \ref{sec:var_ext_coup} allows the exploration of the 250 - 300~MHz range, a new 5 meter long cavity for the 300 - 350 MHz range must be designed. In this way, we would be able to scan a total band of 100~MHz. This new cavity has the same geometry, but a smaller cross-section in order to increase its resonant frequency. \\

\begin{figure}[h]
\centering
\begin{subfigure}[b]{0.7\textwidth}
         \centering
         \includegraphics[width=0.99\textwidth]{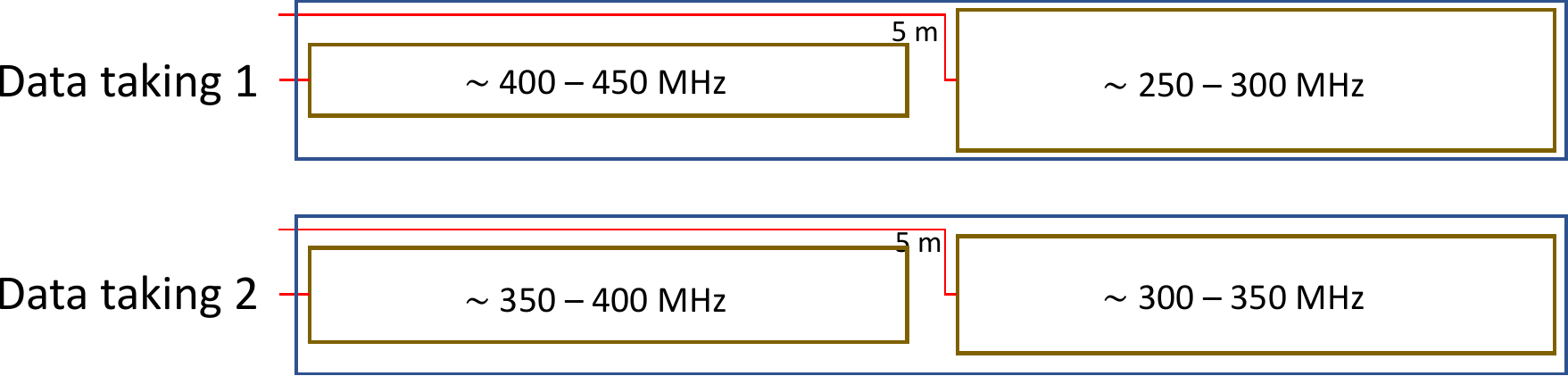}
         \caption{}
          \label{fig:2cav_location}
\end{subfigure}
\hfill
\begin{subfigure}[b]{0.7\textwidth}
         \centering
         \includegraphics[width=0.99\textwidth]{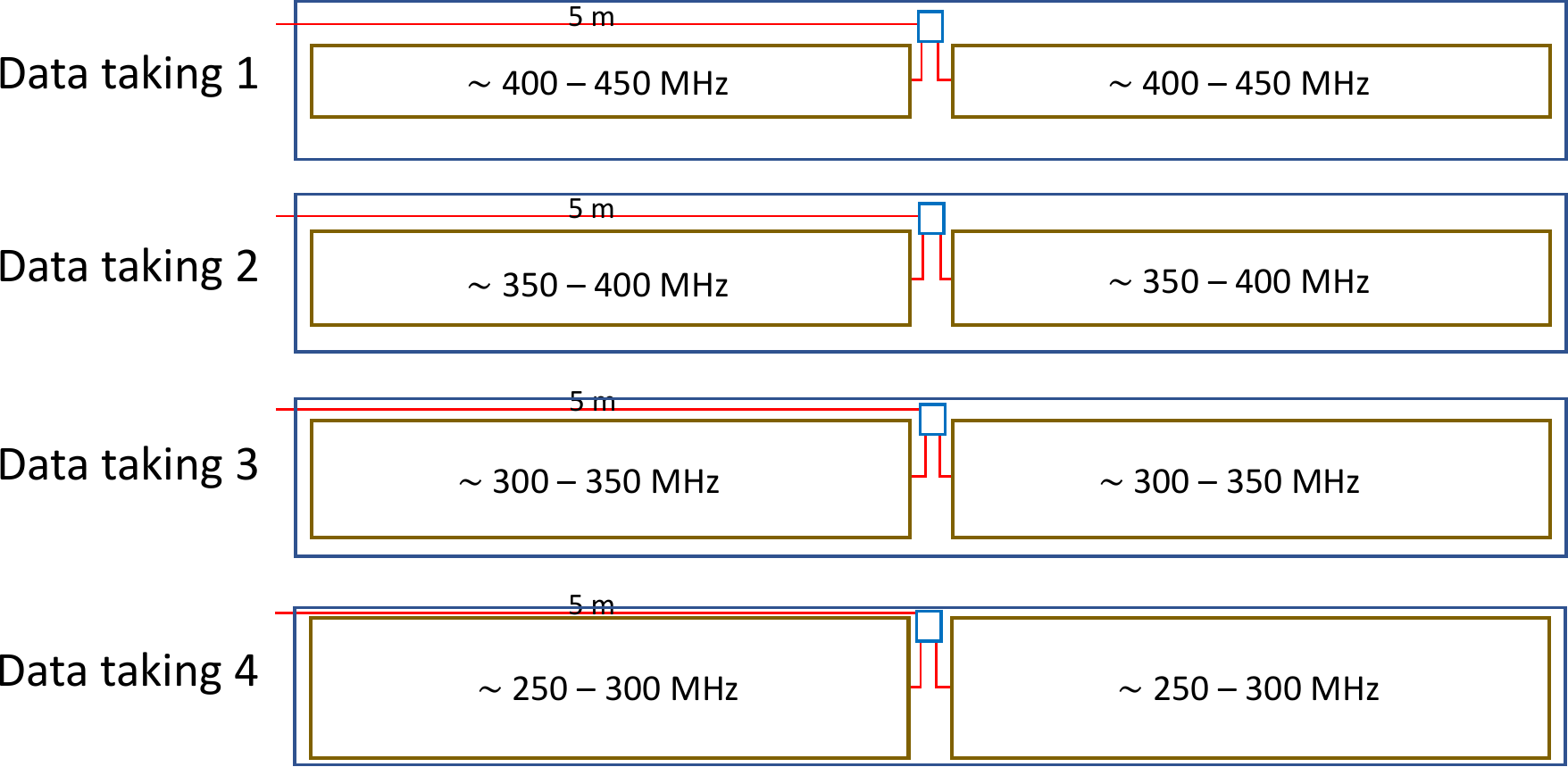}
         \caption{}
         \label{fig:4cav_location}
\end{subfigure}
\caption{Cavities location inside the bore during data taking campaign for (a) parallel mass range exploration, and (b) coherent sum exploration.}
\end{figure}

This concept of multiple mass ranges in the same data taking can be repeated for higher frequencies in order to explore the whole desired range of $1-2$~\textmu eV. For example, the experiment would be composed of two consecutive data-taking campaigns, each one with two different cavities. The location of the cavities inside the bore depends on the mass range covered, in order to compensate the worse sensitivity of higher mass cavities with a lower attenuation due to cable losses (see Figure \ref{fig:2cav_location} for a comparison of the coherent sum procedure of Section \ref{sec:multiple_cav} and Figure \ref{fig:4cav_location} for the parallel exploration alternative).
The total range covered in both cases is 1.046~\textmu eV - 1.941~\textmu eV. Scanning lower masses would require using dielectric material inside the cavity, which would reduce the form factor.

\begin{figure}
  \begin{subfigure}[t]{.49\textwidth}
    \centering
    \includegraphics[width=\linewidth]{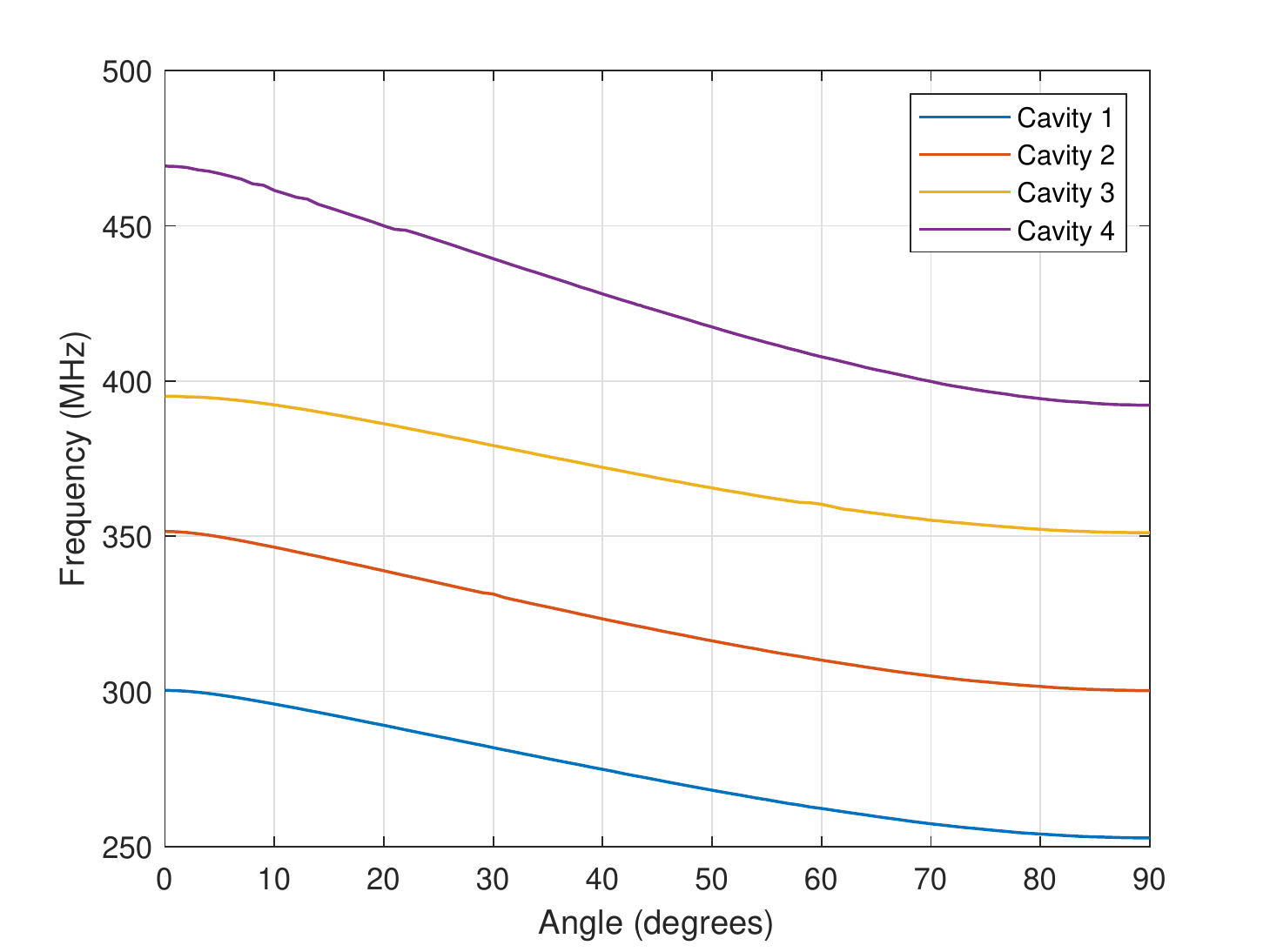}
    \caption{}
    \label{fs_4cav}
  \end{subfigure}
  \hfill
  \begin{subfigure}[t]{.49\textwidth}
    \centering
    \includegraphics[width=\linewidth]{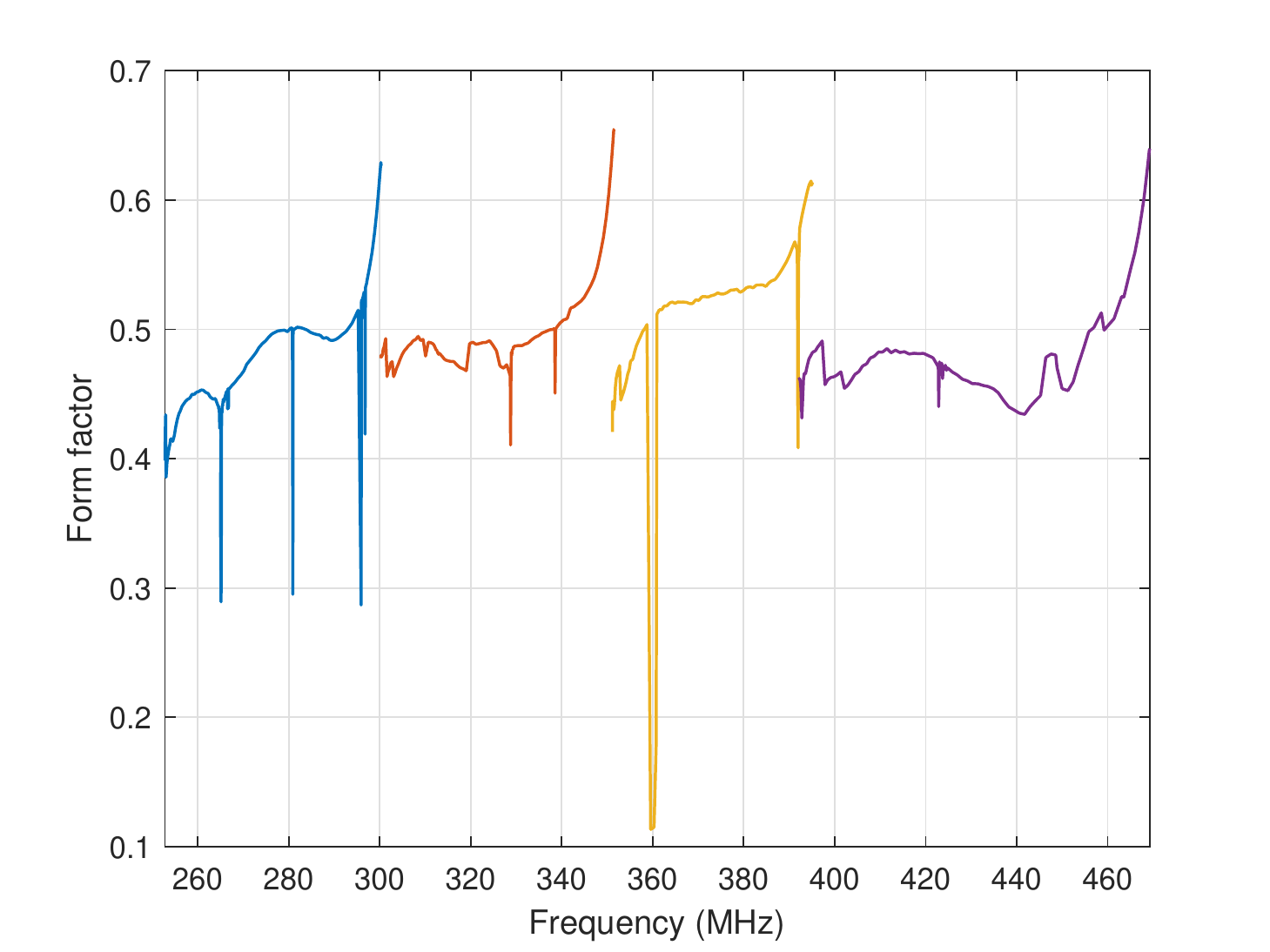}
    \caption{}
    \label{C_4cav}
  \end{subfigure}

  \medskip

  \begin{subfigure}[t]{.49\textwidth}
    \centering
    \includegraphics[width=\linewidth]{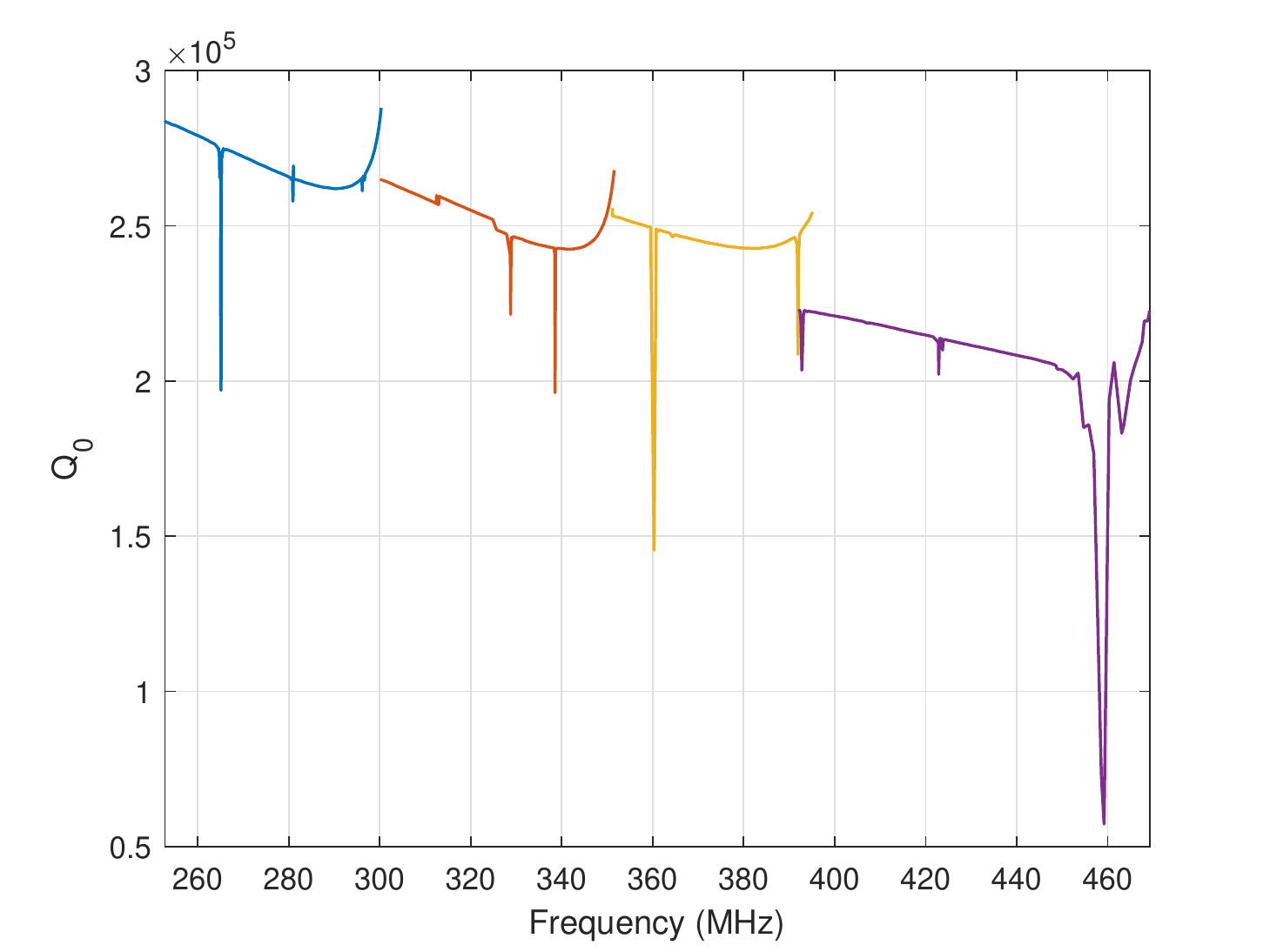}
    \caption{}
    \label{Q0_4cav}
  \end{subfigure}
  \hfill
  \begin{subfigure}[t]{.49\textwidth}
    \centering
    \includegraphics[width=\linewidth]{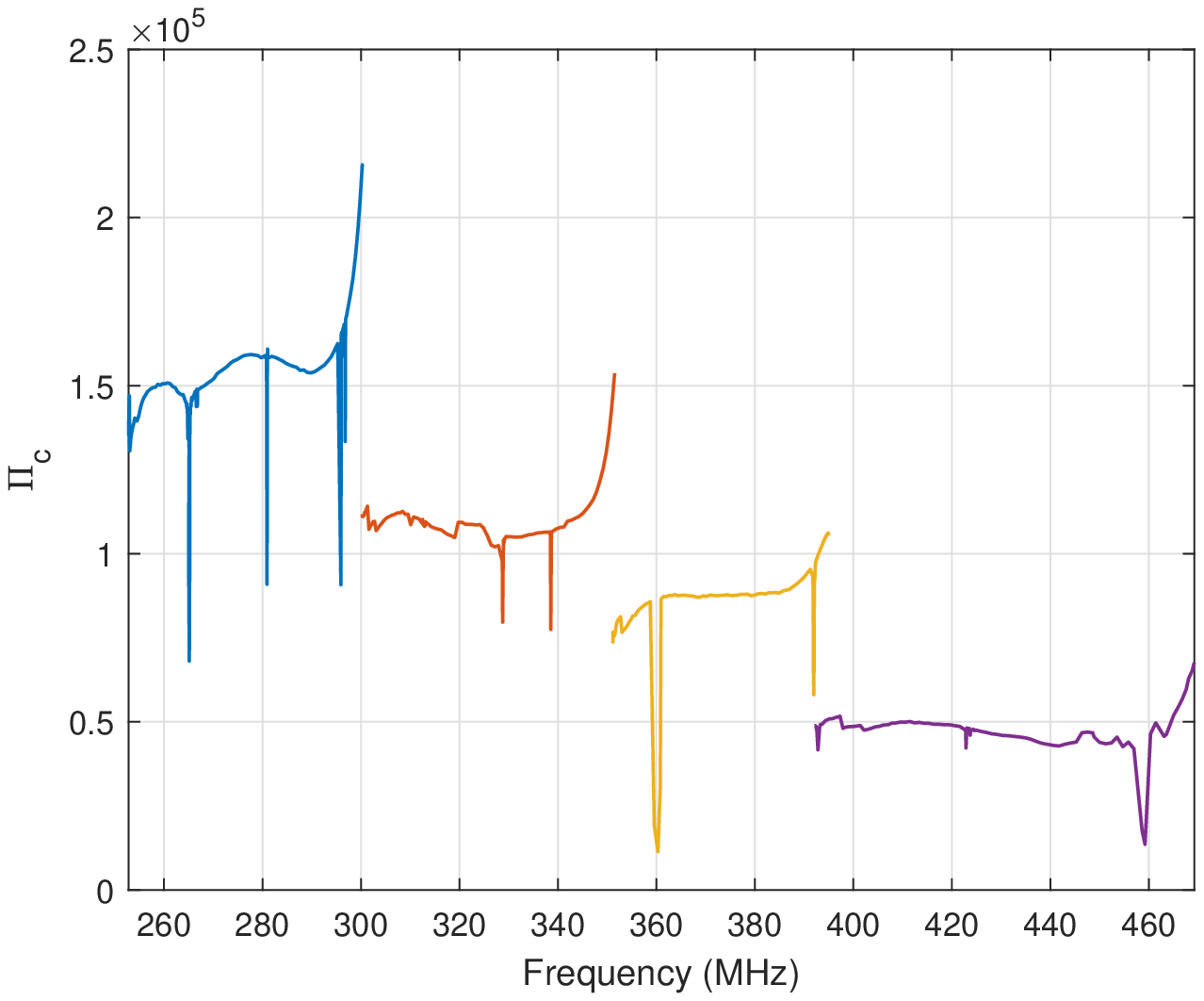}
    \caption{}
    \label{QVC_4cav}
  \end{subfigure}
  \centering
  \caption{(a) Frequency tuning range, (b) Form factor, (c) Unloaded quality factor, and (d) Performance figure $\Pi_c$ for the four designed cavities depending on the frequency.}
\end{figure}

To summarize, Table \ref{tab:cav_dim} shows the main geometrical parameters and tuning range for these four cavities, and Figures \ref{fs_4cav} - \ref{QVC_4cav} show their behavior regarding frequency scanning, form factor, unloaded quality factor and performance figure $\Pi_c$.
\begin{table}[ht]
\small
\centering
\caption{Geometry, volume and tuning range of the designed cavities.
Cavities 1 to 4 cover increasing frequency ranges. The length of all cavities is 5 m. The tuning plates length and thickness are 4.5~m and 10~mm, respectively, their width is $w_p$, and $R_c$ is the cavity corner radius.}
\vspace{2 mm}
\begin{tabular}{c c c c c c c c c c}
\hline
 Cavity & Width (mm) & Height (mm) & $w_p$ (mm) & $R_c$ (mm) &  V ($\textrm{m}^3$) & Tuning range (MHz) \\
\hline
1 & 560 & 504 & 150 & 220 & 1.195 &  252.8 - 300.3\\

2 & 492.3 & 443 & 125 & 220 & 0.876 & 300.3 - 351.6 \\

3 & 430 & 387  & 95 & 180 & 0.682 & 351.2 - 395.2\\

4 & 355 & 319.5 & 95 & 140 &  0.475 & 392.2 - 469.3\\
\hline
\end{tabular}
\label{tab:cav_dim}
\end{table}

\section{Radio frequency (RF) electronics and data acquisition system (DAQ)
\label{sec:rf}
}

\begin{figure}[h!]
\centering
    \begin{subfigure}{0.49\textwidth}
    \includegraphics[width=\textwidth]{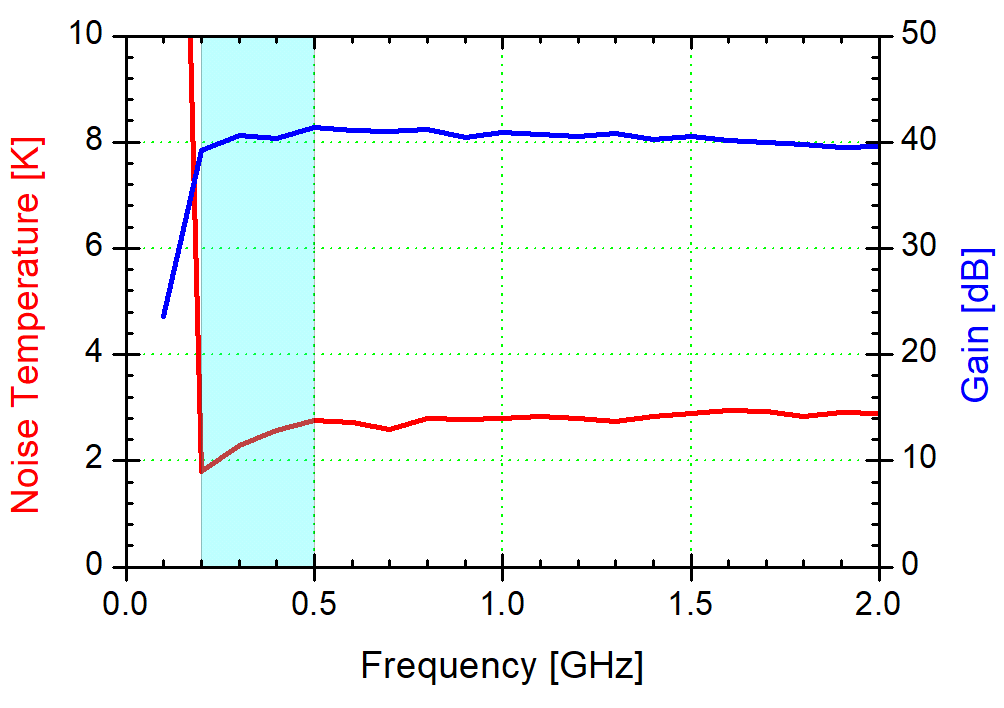}
    \end{subfigure}
    \hfill
    \begin{subfigure}{0.49\textwidth}
    \includegraphics[width=\textwidth]{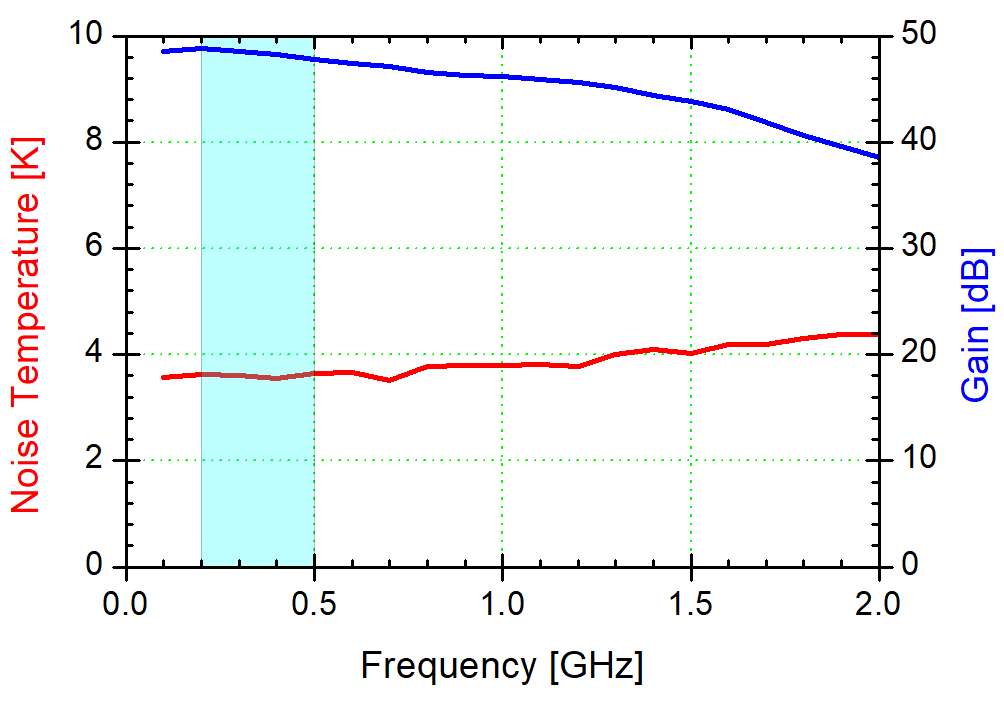}
     \end{subfigure}    
\caption{Measured performance of two practical examples of wide band cryogenic amplifiers usable in the band of interest for BabyIAXO (0.2-0.5 GHz, shadowed in the graph). Left: three stages of InP HEMTs (modified version of ~\cite{InPamplifier}) cooled to 7.5 K. Right: two stages of SiGe HBTs cooled to 8 K ~\cite{SiGeamplifier}. These amplifiers were developed at Observatorio de Yebes for radio astronomy applications.}
\label{fig:InPamp}
\end{figure}

The signal from the cavities must be amplified before being processed. The background noise of the system will be dominated by the physical temperature of the dissipative loss of the cavities and by the noise added by the first stage of amplification. For this reason it is critical to use a good ultra low noise cryogenic amplifier inside the cryostat and to minimize the loss of the coaxial lines between the cavities and the amplifiers. Cryogenic amplifiers for the VHF-UHF frequency range are not as commonly used as those for higher frequency in the microwave range, and obtaining the ultimate noise performance in an octave bandwidth at these frequencies presents some difficulties. Together with the noise, the main problems to address in the amplifier are the stability (in the sense of avoiding auto-oscillations) and the input reflection coefficient.  There are two types of semiconductor device technologies that can be used:
\begin{enumerate}
\item Indium phosphide high electron mobility transistors (InP HEMTs): these devices have been the workhorse for the cryogenic amplifiers used in radio astronomy and deep space communications for many years. They have the potential of obtaining the lowest noise temperatures. However, these devices present very reactive input impedance at low frequency (VHF-UHF), making more difficult obtaining simultaneously low noise and low input reflection. There are some options for procuring devices from this type optimized for low noise cryogenic operation~\cite{Diramics}.
\item Silicon germanium heterojunction bipolar transistors (SiGe HBTs): These devices have the advantage of presenting easier to match input impedance in the VHF-UHF frequency range, although the noise obtained may not be as good as with state of the art InP HEMTs. Their use is less widespread since devices of this type optimized for cryogenic operation are very difficult to procure, but some particular low-cost commercial units have been successfully used obtaining competitive results ~\cite{SiGeamplifier}.
\end{enumerate}
Figure \ref{fig:InPamp} presents the experimental noise and gain results of two practical wideband cryogenic amplifiers built using the two technologies. Note that the measured performance of the amplifiers is referred to an input matched condition (50 Ohm termination). However, in the haloscope, the impedance presented to the input will be that of a critically coupled high Q resonant cavity, which will be very reactive for frequencies out of the resonances. It will be essential to check carefully the unconditional stability of the amplifier (the amplifier should not oscillate when the cavity is connected to its input).  The input reflection is higher in the InP HEMT version than in the SiGe amplifier and this may translate into higher ripples in the output noise power spectrum caused by standing waves in the coaxial line connecting the amplifier to the cavity. The magnitude of this effect should be carefully checked in the final design before taken the decision of the type of amplifier to be used.   

The rest of the signal processing can be carried out at room temperature outside the cryostat with more conventional electronics as in ~\cite{Diaz-Morcillo:2021psa}. First an analog module performs a heterodyne analog conversion of the band to match the working frequency of a digitizer. Once digitized, the FFT is then calculated in real time and integrated to obtain the power spectrum. The digital processing is performed by an FPGA. Such a system was used in RADES for X-band frequency (8.4 GHz) ~\cite{Diaz-Morcillo:2021psa}. In fact, the same hardware developed for RADES X-band can be used with the simple addition of a heterodyne frequency upconverter at the input (200-500 MHz to   8.25-8.55 GHz).

\section{Sensitivity estimates
\label{sec:sensitivity}
}

\subsection{Prospect sensitivity for axions}

Let us now proceed to computing the sensitivity estimate in the axion parameter space for the cavity systems presented in Section \ref{sec:cavity_des}.

We adapt the axion power formula (\ref{eq:Pg}) to BabyIAXO's and standard DM search parameters 

\be
P_d =  7.94 \times 10^{-22}\, {\rm Watt} 
  \frac{\kappa}{0.5} \(\frac{g_{a \gamma}}{2 \times 10^{-16} {\rm GeV}^{-1}}\)^2 
  \frac{1\, \upmu \rm eV}{m_a} \(\frac{B_e}{2\, \rm T}\)^2
  \frac{V}{10^3\, \rm l} \ \frac{Q_l}{3\times 10^5} \
  \frac{C}{0.63}
  \label{signi}
\ee

where $\rho_{\rm DM} = 0.45 {\rm GeV}/ {\rm cm}^3 $ is the value of the dark matter energy density for the local halo.

For a given axion bin width the noise temperature can be obtained as
\be
\label{nosi}
P_T = T_{sys} \Delta \nu_a 
= 6.63\times 10^{-20} \, {\rm Watt}  \frac{T_{sys}}{4.6\, \rm K}  \frac{10^6}{Q_a} \frac{m_a}{1\, \rm \upmu eV} . 
\ee
where we are using the same benchmark values as in Equation \ref{signi}, and $Q_a$ is the natural width of the axion line shape.

We need to know for which scanning time and axion parameters the axion signal power can overcome random noise at each scan step. To evaluate this, we use Dicke's radiometer equation: 
$S/N=P_d/T_{\it sys}  \sqrt{t / \Delta \nu}$. Which combined with \ref{signi} and \ref{nosi} yields
\be
g_{a \gamma}^2 = \frac{S/N \, T_{sys}}{\kappa \rho_{\rm DM} B_e^2 Q_l V C} \( \frac{m_a^3}{t Q_a} \)^\frac{1}{2}
\ee

We want to stress that this expression is defined based on the time for scanning a given mass and the single-frequency SNR, which reflects on a linear dependence on $Q_l$, while in the literature the proportionality quoted is $Q_l^2$ based on the single-frequency SNR squared.

Here, two scenarios are considered: In the first one, two different 5-m long cavities are placed inside the magnet bore at the same time and acquire data with two separated data acquisition systems. These two cavities would be designed to cover slightly different mass regions, and the data taking would last 220 days. Afterwards, these two cavities would be replaced by another pair, again centered around a different mass range and another 220 days of data taking would be performed.
In the second scenario, 4 sets of 2 identical cavities coupled coherently, and connected to a single data acquisition system, are inserted in the magnet for 4 consecutive data takings of 110 days each. The second scenario yields a better result in sensitivity for $g_{a \gamma}$, since it scales with $V^{1/2}$ and $t^{1/4}$.
The potential downside of the second scenario is that the coherent coupling requirement adds a layer of complication to the set-up.

Dividing the total frequency range accessible to our cavity system in scan steps of fixed time (approximately $7 \times 10^4$ steps of 9 minutes each), we find that we can scan at KSVZ sensitivity at almost all the accessible mass range. An exception to this are points in which the performance figure drops in a narrow range (cf. for example Figure \ref{QVC_4cav})

\begin{figure}

\includegraphics[width=\textwidth]{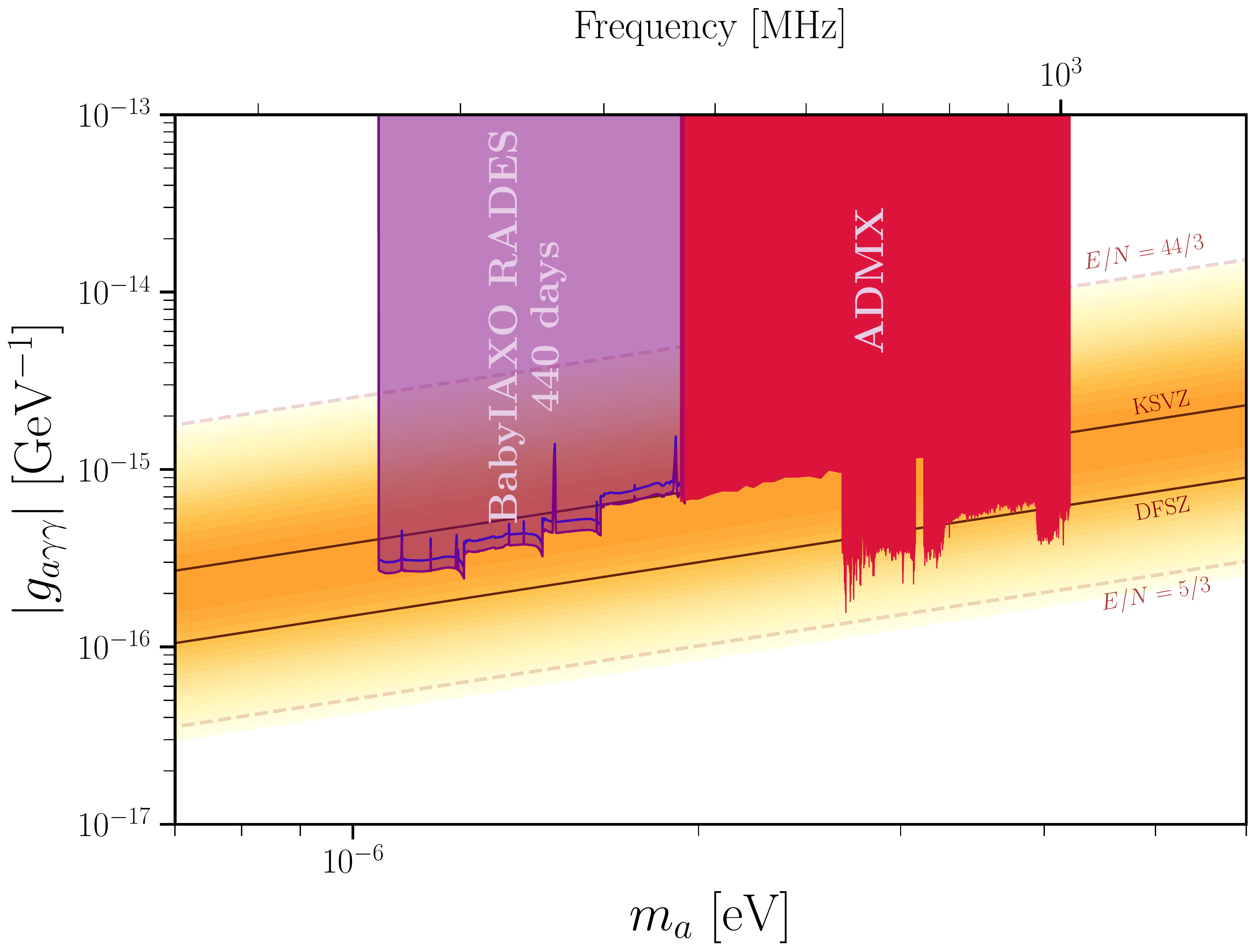}
 \caption{
  Possible BabyIAXO sensitivity (defined as the average 95\% CL upper limit obtained from a background-only data run) using the four 5-m cavities described above, for a total effective exposure time of 440 days. The blue line refers to the scenario defined as parallel acquisition, while the purple one shows the coherent acquisition case. The existing limits from the ADMX experiment are also shown. An adaption of the code in \cite{ciaran} was used for generating the plot.
 }
 \label{fig:babyIaxosens_halo}
\end{figure}

Figure \ref{fig:babyIaxosens_halo} illustrates the results. The two lines in blue and purple correspond to the scenarios described above: blue the independent acquisition for four separate cavities,  purple the coherent coupling for pairs of cavities.

 As can be seen, using the BabyIAXO magnet as a haloscope can probe a parameter space below existing ADMX limits. This parameter space is also targeted by current efforts in ADMX, see \cite{Chakrabarty:2023rha}.

\subsection{Prospect sensitivity for dark photons}

A haloscope experiment for axion detection can also be used to test the existence of massive vector bosons of a $U(1)$ Abelian gauge group symmetry from a sector outside of the SM. This beyond standard model candidates are in general referred to as dark or hidden photons. 
Such a search can even be performed if the external magnet is not on for any reason.
The detection mechanism for such particles is almost equivalent, but some subtleties must be observed, see
\cite{Caputo:2021eaa}: contrary to the axion, the Dark Photon can have a polarization. Thus, depending on assumptions on the polarization of the Dark Matter field, the location of the experiment and its possible relative orientation, different coupling ranges may be probed.

The expected power for DM dark photon conversion on a resonant cavity experiment can be found in \cite{Arias:2012az}:

\be
P_\chi = \kappa  \rchi^2 m_\chi \rho_{\rm DM} V Q_l C_\chi
  \label{sign_DP} 
\ee

where $\rchi$ is the coupling of the dark photon to the SM photon, $m_\chi$ is its mass and the geometric factor $C_\chi$ is computed as
\be
    C_\chi = \frac{\vert\int \vec E \cdot \vec X\: \mathrm d V\vert^2}{\frac{V}{2} \int \epsilon_r \vert\vec E\vert^2\: \mathrm d V}
    \label{eq:form_factor_DP}
\ee
here $\vec X$ is the vector potential for the dark photon, which is ultimately connected to the already mentioned polarization.

For a given mode of our cavity, one can easily see the parallelism between the geometric factors for axions (\ref{eq:form_factor}) and dark photons (\ref{eq:form_factor_DP}). Assuming $\vec B_e$ homogeneous and oriented in the vertical, $\hat y$, direction we can relate both quantities as 
\be
C_\chi = \cos^2(\theta) C
\ee
where $\cos (\theta) = \vec X \cdot \hat y $ is the polarization angle with respect to our setup vertical axis.

\begin{figure}

\includegraphics[width=\textwidth]{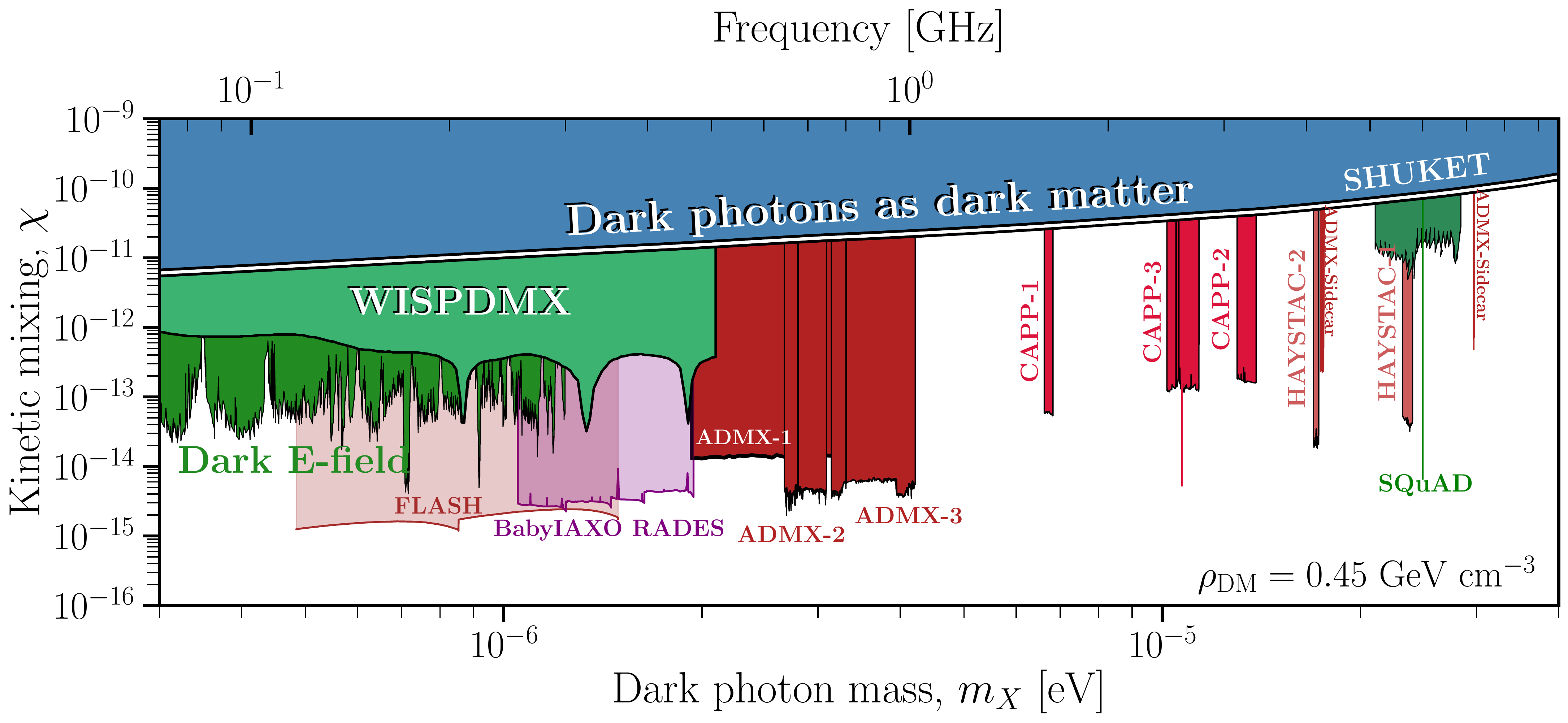}
 \caption{Prospect in sensitivity for a dark photon search performed by a haloscope prototype placed inside of the BabyIAXO telescope, labeled as BabyIAXO RADES on the plot. For this estimation a non polarized cold dark matter halo was assumed. An adaption of the code in \cite{ciaran} was used for generating the plot.}
 \label{fig:babyIaxosens_ciaran_HP}
\end{figure}

When considering dark photon conversion into photons in a haloscope, one has to take into account the experiment movement with respect to the dark matter halo. This, combined with the integration time of the experiment ($T$), will make us not sensitive to a particular value of $\cos^2 (\theta)$ but to an average value $\left\langle \cos^2(\theta) \right\rangle_T$. Let us consider now the fixed polarization scenario for cold dark matter hidden photons, in that case for an experiment with an axis oriented like ours with respect to the axis of rotation of the Earth,  $\left\langle \cos^2(\theta) \right\rangle_T \sim 0.019$ (see \cite{Caputo:2021eaa}). We can then use Equation \ref{sign_DP} to recast the prospects in sensitivity for axions into the equivalent for dark photons, see Figure \ref{fig:babyIaxosens_ciaran_HP}.

\subsection{Prospect sensitivity for HFGWs}

With adaptions to the coupling structures, the BabyIAXO RADES cavities also have sensitivity to hypothesized high-frequency gravitational waves (HFGWs), see, e.g. \cite{Berlin:2021txa,Domcke:2022rgu,Berlin:2023grv}.

We follow the qualitative estimate for the sensitivity to the induced strain $h_0$ of a plane high-frequency gravitational wave of $\sim 2$ min duration in Eq. 29 of \cite{Berlin:2021txa}, with similar benchmark values for the field integral. BabyIAXO RADES achieves in principle sensitivities to strains of $h_0 \sim 10^{-21}$, comparable to ADMX.
Note that accounting for the effect of mechanical deformations (mechanical bars) can lead to an even higher sensitivity \cite{Berlin:2023grv}.

More detailed studies in this direction are currently ongoing in our group.

\section{Alternative detector concepts at lower masses
\label{sec:BASE-DM}
}

To probe axion masses lower than $\sim$1 \textmu eV an alternative haloscope concept, the lumped element LC circuit, has been proposed by Sikivie, Sulivan and Tanner \cite{SKT}, and adopted by collaborations including ADMX-SLIC \cite{ADMXSLIC}, DM-Radio \cite{DMRadio:2022pkf, DMRadio2, DMRadio3}, ABRACADABRA (now part of DM-Radio) \cite{ABRA}, SHAFT \cite{SHAFT}, BASE-CDM \cite{devlin} and others \cite{J.Kim, Zhang}, although other non-LC detector concepts \cite{BEAST, Berlin} are also under development. The LC haloscope is a suitable way to detect axion masses with a corresponding Compton wavelength $\lambda_a=h/(m_a c)$ much larger than any of the dimensions of the haloscope. In an external static magnetic field $\textbf{B}_e$, and neglecting spatial variations of the axion-field, a time dependent magnetic field $\textbf{B}_a(t)$ is generated by the effective electric current density $\textbf{j}_a=-g_{a\gamma}\textbf{B}_e\overset{.}{a}$ arising from the conversion of dark matter axions. Using appropriate detector geometries, sensitive superconducting LC circuits are suitable to detect these oscillating magnetic fields $\textbf{B}_a(t)$. It is interesting to consider what hypothetical limits could be set on lower mass axions by placing such an LC haloscope inside the BabyIAXO magnet to motivate future, more detailed technical work in this direction. In the following sections, we consider several coil geometries and approaches. We start by directly adapting the analysis presented in Ref. \cite{SKT} to the dimensions, temperatures and field strength of BabyIAXO. This is helpful to compare the BabyIAXO magnet to other magnets discussed in Ref. \cite{SKT} and to the recent ADMX-SLIC results \cite{ADMXSLIC}, which use this approach. In subsequent sections, we consider more optimized coil geometries and more efficient scanning strategies suitable for lower mass axions.

\subsection{Single rectangular loop}
\label{section:singleTurnLoop}

\begin{figure}[h]
\centering
\includegraphics[width=0.9\textwidth]{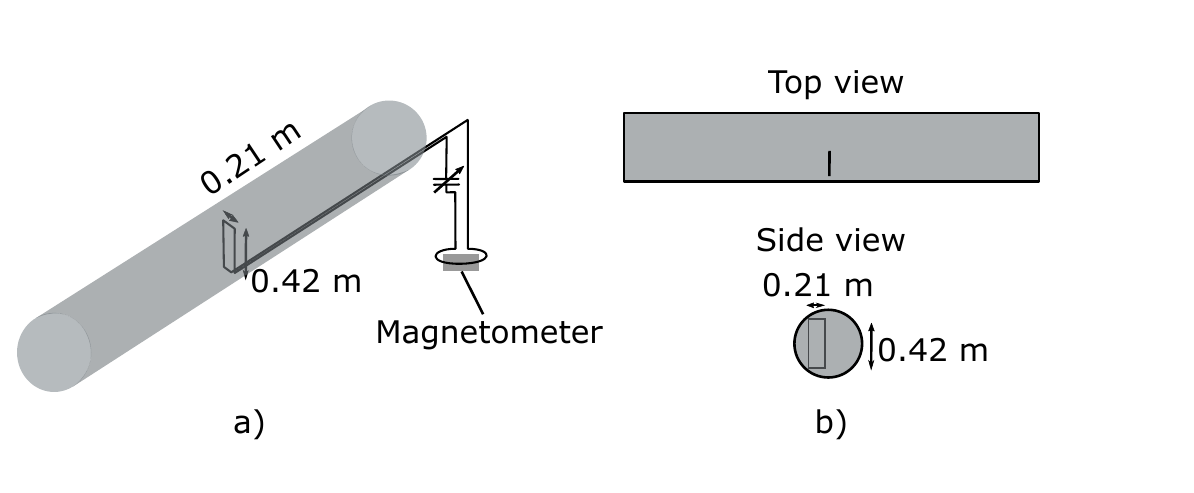}
\caption{a) Illustration of the LC detection geometry with a single pickup loop b) Top view and side view of the single loop geometry}
\label{fig:single_loop}
\end{figure}

We begin with the case first considered in Ref. \cite{SKT} illustrated in Figure \ref{fig:single_loop} a) and b). This consists of a single pickup-loop inside the magnetic field, connected in series with an adjustable capacitance and a secondary detection coil. Any current induced in the primary coil drives the secondary coil, which in turn produces a magnetic field which can be detected by a sensitive magnetometer coupled to the secondary coil. We consider a rectangular loop with horizontal dimension $r_m=0.21$~m and vertical dimension $l_m=0.42$~m centrally located along the axis of the cryostat bore and offset radially so it occupies the left-hand side of the bore, as shown in Figure \ref{fig:single_loop}.

The current induced in the detection coil is given by 
\begin{align}
I&=\frac{Q}{L\omega}\frac{d}{d t}\iint_\Sigma\textbf{B}_a(t)\cdot \textbf{dA}\nonumber \\
&= 4.16 \times 10^{-12}\textrm{A}\(\frac{Q}{10^4}\)\(\frac{\upmu H}{L}\)\(\frac{V_m}{\textrm{m}^3}\)\(\frac{g_{a\gamma}}{10^{-17}\textrm{ GeV}^{-1}}\)\(\frac{\rho_a}{\textrm{0.45~GeV}/\textrm{cm}^3}\)^{\frac{1}{2}}\(\frac{B_e}{2 \textrm{T}}\).
\label{eq:singleLoopCurrent}
\end{align}
Here $Q$ is the quality factor of the circuit, ${V_m=\frac{1}{g_{a\gamma}B_e\overset{.}{a}}\int\int_\Sigma\textbf{B}_a(t)\cdot \textbf{dA}}$ is a geometrical factor proportional to the integrated flux, $\omega$ is the resonant frequency, $L$ is the total circuit inductance and $B_e=|\textbf{B}_e|$ the external magnetic field that is aligned parallel to the long side of the pickup loop. In the second line, we have replaced the axion field amplitude with the local dark matter density $\rho_a$, assuming that the entire dark matter density is represented by the field $a$. A numerical evaluation of $V_m$ using COMSOL, assuming a homogenous $\textbf{B}_e$ and the stated coil dimensions, yields $V_m= 0.0093$ m$^3$ . To compute the expected limits, we compare the axion signal to the noise sources of the detector. In principle, these include both the technical noise floor of the magnetometer and the Johnson noise current of the circuit. We imagine a high-Q circuit ($Q=10^4$, as in ADMX-SLIC), where we only consider axion signals within the full-width-at-half-maximum around the resonant frequency of the circuit, and we consider a magnetometer with a noise floor below 0.1$\,$fT$/\sqrt{\text{Hz}}$, feasible with optimized atomic rf magnetometer designs \cite{J.Kim} up to ~10 MHz. In this case, the technical noise of the magnetometer can be neglected, and the noise is given by the Johnson noise current 
\begin{align}
    \delta I_T=\sqrt{\frac{4k_BTQ\Delta\nu}{L\omega}}=2.96 \times 10^{-10} \textrm{A} \(\frac{\textrm{MHz}}{\nu}\)^{\frac{1}{2}}\(\frac{\upmu\textrm{H}}{L}\)^{\frac{1}{2}}\(\frac{Q}{10^4}\)^{\frac{1}{2}}\(\frac{T}{\textrm{K}}\)^{\frac{1}{2}}\(\frac{\Delta\nu}{\textrm{Hz}}\)^{\frac{1}{2}}.
    \label{eq:singleLoopNoise}
\end{align}
We follow the assumptions of Ref. \cite{SKT} and conservatively set an exclusion limit where the signal $I$ is more than five times the noise $I>5 \delta I_B$. We further require that a factor of two in frequency should be scanned each year at a duty cycle of 30\%. For easy comparison we adopt the conservative assumption that the detector should be stepped in steps of its bandwidth $\nu_R/Q$, in reality, as we will see later, it will likely be more efficient to take larger steps and spend more time at each frequency tuning \cite{Chaudhuri2018}. In this case we allow $10^3$ seconds at each capacitor point. The total inductance $L=L_m+L_d$ is the sum of the coil inductance $L_m$  calculated using $L_m=\frac{\mu_0}{2\pi}l_m\textrm{ln}(\frac{r_m}{a}) =1$ $\upmu$H, where $a=4 \upmu$m is the wire thickness, and a fixed inductance $L_d=0.2$ $\upmu$H to account for the inductance of the leads. We use $Q=10^4$, $r_m=0.21$ m, $l_m=0.42$ m, $T=4.6$ K, $B_e=2$ T and assume $\rho_a=0.45$ GeV/cm$^3$. The circuit has some minimum capacitance $C$ which limits the maximum resonant frequency to which it can be tuned via $2\pi\nu_R=(LC)^{-\frac{1}{2}}$, to calculate this we used a capacitance of 15 pF per meter of circuit. The resulting limits are shown in the light purple region in Figure \ref{fig:LC_limits} labelled ``i''. We note the the projected limits are similar to the narrow band limits that the ADMX-SLIC experiment obtained with a similar sized coil in a 7 T magnet, albeit at $\sim$ 3x longer averaging time per point compared to this proposal.

\subsection{Toroidal Coils}

To improve on the low frequency haloscope design, we now consider an optimized pickup geometry to extract the best limits on the axion-photon coupling given a particular target frequency and available magnetic volume. To tackle this problem, we restrict ourselves to considering toroidal coils with rectangular cross sections, such that the largest possible coil we could imagine has a height and diameter of 0.42$\,$m, so that it fits inside the bore of the BabyIAXO magnet.  Building on devices developed for cryogenic detection of single particle image currents \cite{Nagahama}, the BASE-CDM experiment recently demonstrated a small axion haloscope based on a $\sim$1100 turn high-$Q$ toroidal coil with outer diameter 38$\,$mm and height 22$\,$mm \cite{devlin}. We imagine placing much larger versions and many of these coils with various different numbers of winding inside the BabyIAXO magnet. 

\begin{figure}[h]
\centering
\includegraphics[width=0.5\textwidth]{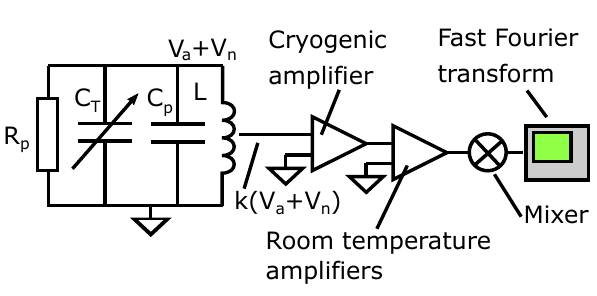}
\caption{The effective circuit diagram of the resonant coil haloscope. The detection inductor is represented by the elements $R_p$, $C_p$ and L, in parallel is an adjustable capacitor $C_T$ to tune the sensitive frequency band. The signal is picked-up by an ultra-low-noise cryogenic amplifier, post-amplified and processed with mixer stages and Fast Fourier Transform spectrum analyzers. }
\label{fig:LC_BASECOIL}
\end{figure}

For the resonant coils, the comparatively large parasitic capacitance across the coil inductance necessitates a variation of the detection strategy presented above. In this case the adjustable capacitance is placed in parallel with the resonant coil, and a voltage rather than a current is detected. The effective circuit diagram of this arrangement is illustrated in Figure \ref{fig:LC_BASECOIL}. Here $L$ is the total inductance of the  coil, $C_p$ is the parasitic parallel capacitance of the circuit, $C_T$ is the tunable capacitance and $R_p$ is the parallel resistance. In this configuration, the axion-sourced magnetic field $\textbf{B}_a$ leads to an electromotive force across the coil. A fraction $\kappa$ of the voltage is picked-up by means of a tapped transformer connection at the grounded end of the coil, and this is subsequently connected to a cryogenic low-noise, high input-impedance GaAs amplifier \cite{Nagahama, Ulmer}. The tunable capacitor can be realized using moving copper plates; the BASE-CDM project has realized 58 pF of capacitance tuning with a copper parallel plate capacitor \cite{voelksen} on a background capacitance of 10$\,$pF, and is also developing a rotary capacitor with $>$150$\,$pF tuning range.     

Figure \ref{fig:LC_BASEDetectorConcept}  gives a 3/4 cut view of the single detection toroid under consideration. The bulk of the volume contains an N-turn rectangular section toroidal inductor with  height $l_m$, inner radius $r_1$ and outer radius $r_2$. The inductor will be made from type-II superconducting NbTi wire or narrow high TC superconducting tape, and placed in a high-TC superconducting coated housing or a NbTi housing, which will define the capacitance, and shield the coil from rf-pickup and lossy elements. Also shown are the adjustable rotary capacitor and the cryogenic amplifier.

\textbf{\begin{figure}[h]
\centering
\includegraphics[width=0.9\textwidth]{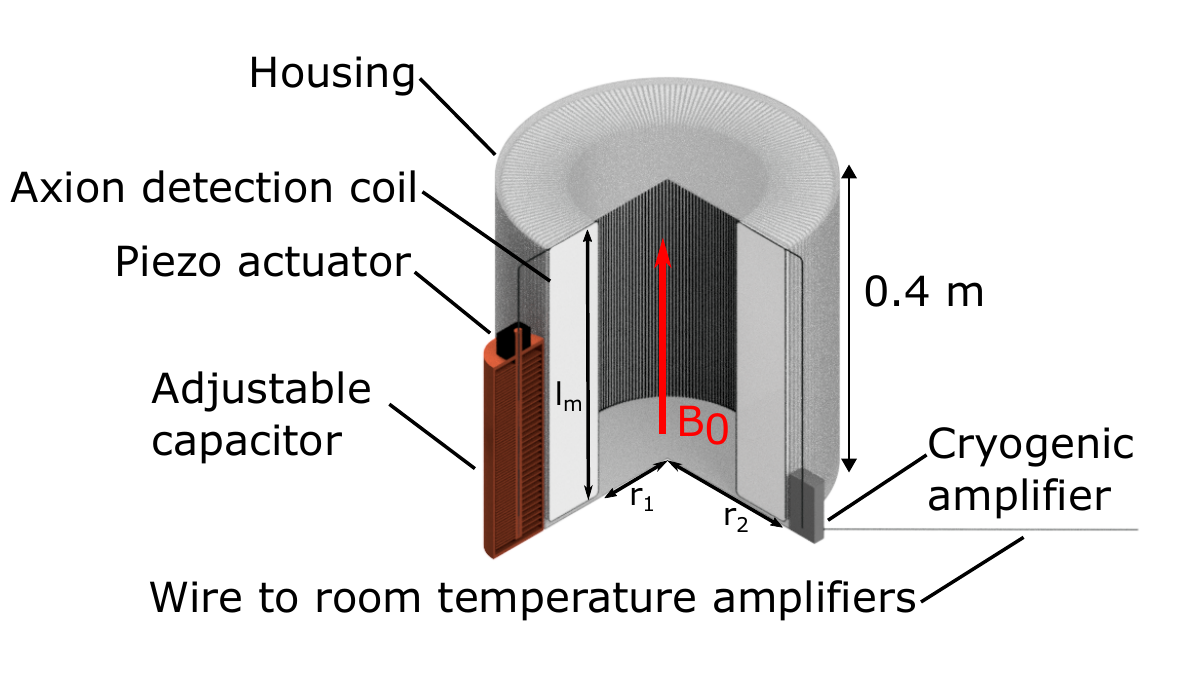}
\caption{Layout of a detection toroid to be placed in the BabyIAXO magnet bore.}
\label{fig:LC_BASEDetectorConcept}
\end{figure} }

We now consider the performance of a coil like that shown in Figure \ref{fig:LC_BASEDetectorConcept}. The total power at the input of the cryogenic amplifier can be written as the sum of the signal power $\propto V_a^2$ and noise power $\propto V_n^2$ which itself is the sum of Johnson noise and the amplifier's equivalent input noise. The ratio of the signal power to the noise power fluctuations $V_a^2/\sigma(V_n^2)$ is given by \cite{devlin}

\begin{align}
\frac{V_a^2}{\sigma(V_n^2)}&=g_{a\gamma}^2 K S \sqrt{\frac{Q_a}{\nu_a}}\sqrt{ t_\textrm{av}}\/.,
\label{EQ:LC_axion_basis}
\end{align}

\noindent where $\nu_a/Q_a$ is the frequency bandwidth of the axion signal with frequency $\nu_a$, $t_\textrm{av}$ is the averaging time and the term $K$ is

\begin{align}
K &=\frac{\nu_a Q\pi \rho_a \hbar c (l_m N(r_2^2-r_1^2)B_0)^2 }{32 L k_B T},
\end{align}

\noindent when the haloscope is at a temperature $T$. The term $S$ corrects for the reduction in the signal-to-noise ratio if the axion signal is not exactly at the peak of the LC circuit resonance. It can be expressed as   
\begin{align}
S &=\frac{1}{1+f^{-1} D_\textrm{SNR}^{-2}}.
\end{align}
Here $f$ is a Lorentzian lineshape function normalized to have unit height if the axion frequency matches the detector's resonant frequency $\nu_R$\;
\begin{align}
\frac{1}{f} &=1+\frac{4Q^2(\nu_a-\nu_R)^2}{\nu_R^2},
\end{align}
and $D_\textrm{SNR}=\frac{\kappa \sqrt{8 \pi k_B T \nu_R Q L}}{en}$ is the ratio of the Johnson noise of the circuit at the resonant frequency and the equivalent input noise voltage $e_n$ of the cryogenic amplifier. We also refer to $D_\textrm{SNR}$ as the detector signal-to-noise ratio. When $\nu_a$ matches $\nu_R$ then $S=\frac{D_\textrm{SNR}^2}{1+D_\textrm{SNR}^2}\simeq1$. 

We can use Equation \ref{EQ:LC_axion_basis} to predict the performance of the haloscope in various situations. A given axion signal would be expected to exceed the local 5-$\sigma$ threshold for discovery if the right hand side exceeds $V_a^2/\sigma(V_n^2)=5$. On the other hand, if we would like to predict where the 95$\,$\% confidence limits to exclude axions with a certain mass and coupling strength would be expected, we can solve Equation \ref{EQ:LC_axion_basis} for $V_a^2/\sigma(V_n^2)=2$. If the axion signal appears on resonance, then the resulting limit, setting $Q_a=10^{6}$, is given by
\begin{align}
    g_{a\gamma}(\textrm{resonant})=3.88 \times 10^{-12} \,\, \textrm{GeV}^{-1}
    &\(\frac{\textrm{m}^3 }{N l_m (r_2^2-r_1^2)}\) 
    \(\frac{2\,\,\textrm{T}}{B_e}\)
    \(\frac{10^4}{Q}\)^{\frac{1}{2}}
    \(\frac{0.45\,\,\textrm{GeV/cm}^3}{\rho_a}\)^{\frac{1}{2}}\nonumber\\
       &
    \times \(\frac{L}{\textrm{mH}}\)^{\frac{1}{2}}\( \frac{T}{4.6 K}\)^{\frac{1}{2}}
    \(\frac{10^3 \,\,\textrm{s}}{t_\textrm{av}}\)^{\frac{1}{4}}
    \(\frac{\textrm{MHz}}{\nu_a}\)^{\frac{1}{4}}.
    \label{eq:LC_axion_resonant_Limit}
\end{align}

Equation \ref{eq:LC_axion_resonant_Limit} is useful for assessing the optimal performance of the haloscope for axion signals which match the resonant frequency of the circuit. When the detector's resonant frequency is scanned in steps $\Delta f$, the upper limit on $g_{a\gamma}$ that is achieved is that which is set at the edge of the scanning bandwidth for each scan, given by  $g_{a\gamma}(\textrm{scanning})=S^{-1/2}g_{a\gamma}(\textrm{resonant})$. With a total scanning time $t_\textrm{tot}=N_st_\textrm{av}$ and scanning between $f_\textrm{max}$ and $f_\textrm{min}$, the limit $g_{a\gamma}(\textrm{scanning})\propto S^{-1/2}\sqrt{N_s/t_\textrm{tot}}\propto(\Delta f S )^{-\frac{1}{2}}$, so the optimal step will be that which maximises $\Delta f S$. This step is $\Delta f=\nu_R\sqrt{1+D_\textrm{SNR}^2}/Q$. This is approximately the detector bandwidth $\nu_R/Q$ multiplied by the SNR, which implies significantly larger steps than in the initial analysis. When performing this optimal step, the limits at the edge of the detection bandwidth would be expected to be 1.11$\times$ worse than at the middle. Substituting in these optimum parameters, we find that the expected limit for this scanning experiment is given by

\begin{align}
    g_{a\gamma}(\textrm{scanning})=3.88\times 10^{-14} \,\, \textrm{GeV}^{-1}
    &\(\frac{\textrm{m}^3 }{N l_m (r_2^2-r_1^2)}\) 
    \(\frac{2\,\,\textrm{T}}{B_e}\)
    \(\frac{0.45\,\,\textrm{GeV/cm}^3}{\rho_a}\)^{\frac{1}{2}}\nonumber\\
       &
    \times \(\frac{L}{\textrm{mH}}\)^{\frac{3}{8}}\( \frac{T}{4.6 \,\, \textrm{K}}\)^{\frac{3}{8}}
    \(\frac{10^4}{Q}\)^{\frac{3}{8}}
    \(\frac{\textrm{MHz}}{\nu_a}\)^{\frac{3}{8}}
    \nonumber\\
       &
       \times
    \(\frac{\textrm{year}}{t_\textrm{tot}}\)^{\frac{1}{4}}
    \(\frac{\textrm{ln}(2)}{\textrm{ln}(f_\textrm{max}/f_\textrm{min})}\)^{\frac{1}{4}}
     \(\frac{e_n/\kappa}{10 \textrm{nV}/\sqrt{\textrm{Hz}}}\)^{\frac{1}{4}}.
    \label{eq:LC_axion_scanning_Limit}
\end{align}

With this expression, it is possible to optimize the geometry of the coil to set the most stringent limits. For simplicity, we assume that the quality factor is independent of the frequency and dimensions of the coil. This is well justified if the predominant losses in the circuit are due to the finite input resistance of the cryogenic amplifier. In this case, to design the optimum coil at a particular frequency $\nu_R$ we simple need to minimize $\(\frac{L }{N l_m (r_2^2-r_1^2)}\)$ while fixing the product $(C_p L)^{-1/2}=2\pi \nu_R$. We calculate the coil inductance using $L=\frac{\mu_0}{2\pi}N^2l_m\textrm{ln}(\frac{r_2}{r_1})$ while the parasitic capacitance can be determined considering the individual capacitance between neighboring coils and between the coils and the housing \cite{capacitanceCalc}. We fix the spacing between the coil and the housing at 5 mm and choose $l_m=2r_2=0.4$ m. We choose a wire diameter of 75 $\mu$m with and outer PTFE coating of thickness 22.5 $\mu$m. The inner radius and number of turns are optimized to set the lowest limit on $g_{a\gamma}$. For the finalized detector, we envision an array of 11 separate coils which can be operated simultaneously. This concept is shown in Figure \ref{fig:LC_BASEDetectorsInBore}. We allocate each coil to scan a factor 2 in frequency, with the upper coil probing a maximum frequency of 40 MHz. To compute the limits we assume a 1-year scan time with a measurement duty cycle of 30\%, and other fixed parameters are as in Equation \ref{eq:LC_axion_scanning_Limit}.

\textbf{\begin{figure}[h]
\centering
\includegraphics[width=0.9\textwidth]{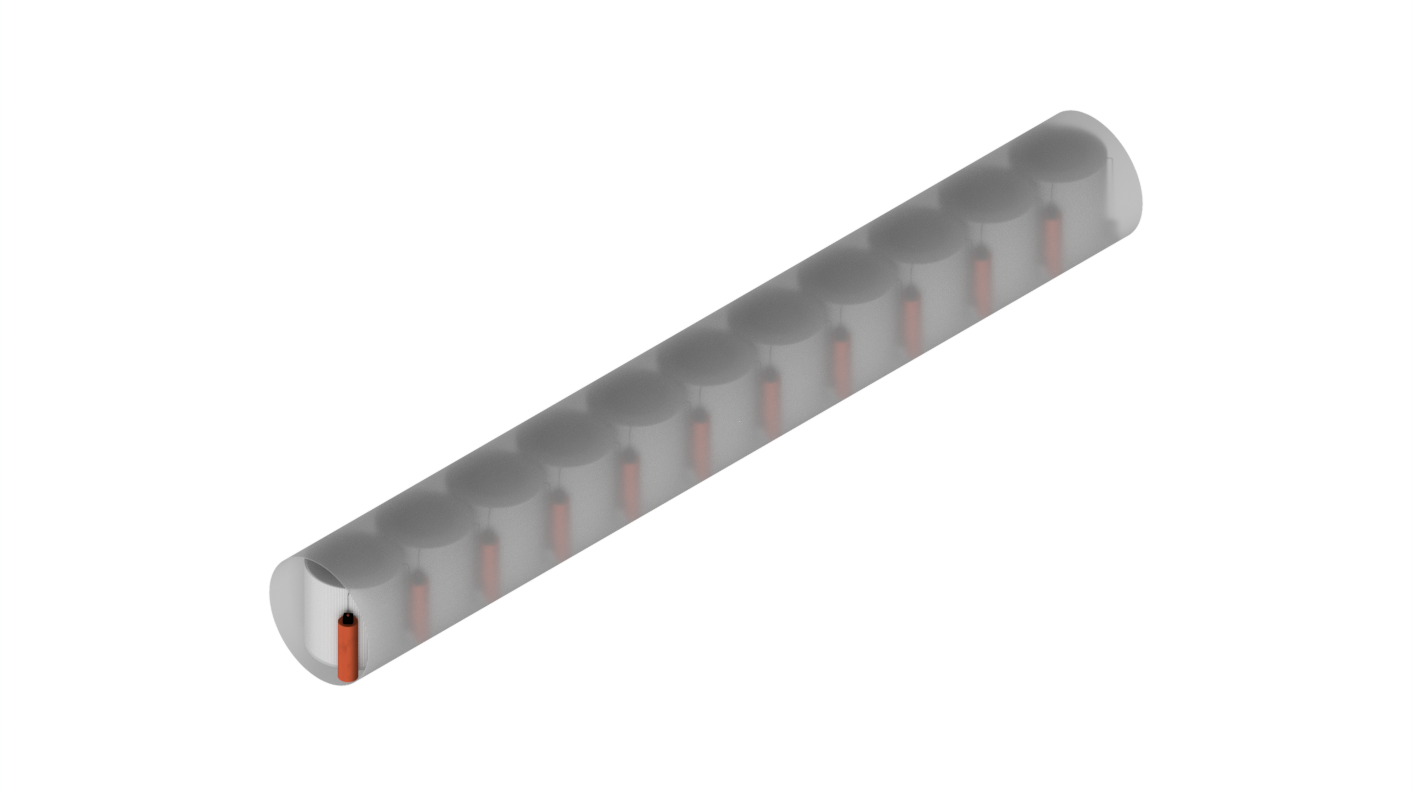}
\caption{A series of 11 haloscopes placed in the bore with different numbers of windings, each designed to probe a different resonant frequency}
\label{fig:LC_BASEDetectorsInBore}
\end{figure} }

\begin{figure}

\includegraphics[width=\textwidth]{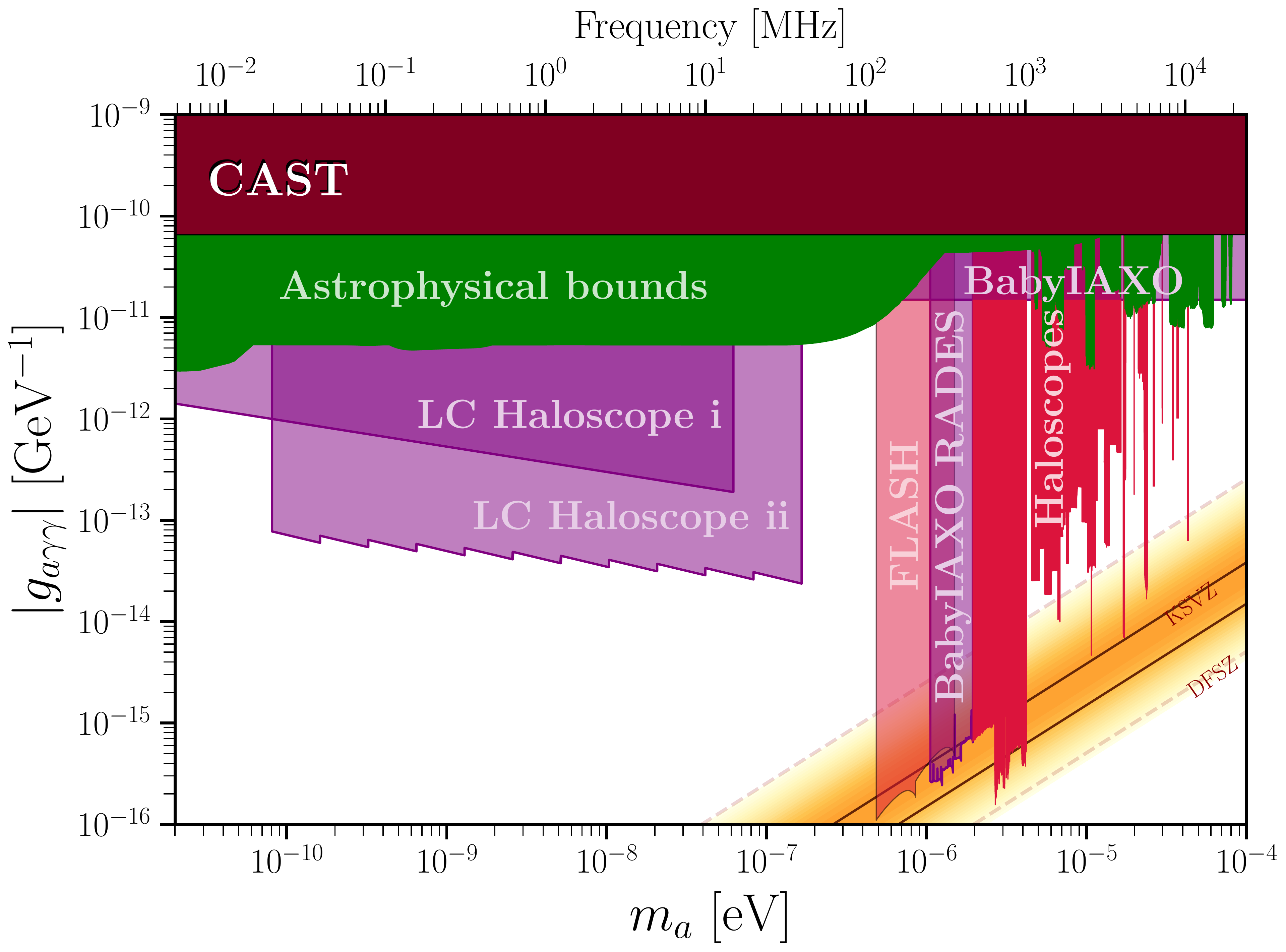}
 \caption{Prospects in sensitivity in the axion mass vs axion-photon coupling parameter space for LC haloscope searches at BabyIAXO. LC Haloscope \textit{i} refers to acquisition with a rectangular coil, while LC Haloscope \textit{ii} refers to acquisition with a toroidal coil. The prospects for the BabyIAXO RADES setup discussed in Section \ref{sec:sensitivity} and a part of the BabyIAXO sensitivy are also displayed. The limits projected by the FLASH experiment are also shown in salmon \cite{Flash}. Relevant excluded areas in this region are shown, i.e. limits from CAST, haloscope experiments, astrophysical bounds and axion DM conversion in neutron stars. An adaption of the code in \cite{ciaran} was used for generating the plot, an exhaustive list of references can be found in the repository. }
 \label{fig:LC_limits}
\end{figure}

The results of the optimization are plotted in light purple in Figure \ref{fig:LC_limits}, and marked ``ii''. We see that this more sophisticated approach allows us to set limits around a factor of 5 better than the previous approach, in a total scan time of 1 year rather than 11 years as required to achieve the single limits marked i) with a single coil. This can be allocated to a number of factors: using the 2-$\sigma$ exclusion limits rather than the 5-$\sigma$ discovery threshold (factor 1.6), using coils with lower inductance and higher sensitive volume (factor 1.25), and the optimized scanning strategy (factor 2.3). 

It is likely that the design of the coils could be further optimized by extending the coils into the space above and below the housing and to the edges of the magnet bore. It may also be possible to realize higher quality factors than the assumed $10^4$, recently  BASE-CDM demonstrated a coil operating at 500-720 kHz with a $Q>90,000$ showing that there is scope to improve. There is also modest scope to reduce the coupled equivalent input noise $e_n/\kappa$. The listed figure in Equation \ref{eq:LC_axion_scanning_Limit} is typically realized in the BASE detectors, where it has been optimized to a level where it no longer imposes any limitations on the experiment. For this application we could either increase the coupling factor $\kappa$ or use transistors with lower $e_n$ such as GaAs PHEMT's \cite{betterTransistor}. We think that this approach can set interesting limits in the lower mass axion space and is sufficiently well motivated to warrant further investigation. 

\section{Conclusion
\label{sec:conc}
}

The proposed 
IAXO helioscope and its predecessor, the BabyIAXO experiment \cite{IAXO:2020wwp} will be the ultimate realistically conceivable axion helioscope. Both experiments are aiming to find or exclude  axions from the Sun. IAXO will not only be unprecedented in its sensitivity to axions over a vast mass reach, but will also be one of the few axion experiments to receive its own powerful purpose-made superconducting magnet.

It is only natural to ask how the impact on axion searches of such a magnet can be optimized in every respect over its lifetime. For this reason, and - more importantly - due to the theoretical attractiveness of axions in the 100-500 MHz range, in this article we have put forward ideas and strategies for the exploitation of the power of the BabyIAXO magnet also as an axion haloscope.
Together with the proposed FLASH haloscope, a large region below the current ADMX mass range could be explored over the next decade, see \cite{workshop} for a recent overview of ideas.

One of the novelties of this proposal is that radio-frequency (RF) cavities suited for the purpose have to be conceived for a magnetic {\it dipole}, contrary to most haloscope experiments which utilize {\it solenoid} magnets.  
With this in mind, studies have been initiated within the Relic Axion Dark Matter Exploratory Search (RADES) project  in the past years.
The RADES group built several long coupled-like cavities, which were successfully used in the last years of the CAST magnet operation at CERN, as summarized in a recent overview article \cite{Diaz-Morcillo:2021psa}.

In this paper we have extended this previous research with prototypes in the tens of GHz range to the scale of the full BabyIAXO magnet bore at a few hundred MHz.
We have also described how an LC circuit-based set-up could be implemented in BabyIAXO to probe ALP parameter space from below the 100 neV scale.

In view of future haloscope measurements at BabyIAXO, the RADES group is planning next to take measurements with prototype cavities that scrutinize the concepts presented in this article.
Given the full-scale experiment prospect sensitivity to axions, dark photons and even high-frequency gravitational waves, utilizing BabyIAXO not only as a helioscope but also as a haloscope seems a very attractive opportunity.

\section*{Acknowledgements}

This work has been funded by MCIN/AEI/10.13039/501100011033/ and by ``ERDF A way of making Europe'', under grants PID2019-108122GB-C33 and PGC2022-126078NB-C21. We acknowledge Grant  DGA-FSE grant 2020-E21-17R Aragon Government and the European
Union - NextGenerationEU Recovery and Resilience Program on ``Astrofísica y Física de Altas Energías'' CEFCA-CAPA-ITAINNOVA. This article/publication is based upon work from COST Action COSMIC WISPers CA21106, supported by COST (European Cooperation in Science and Technology). JMGB thanks the grant FPI BES-2017-079787, funded by MCIN/AEI/10.13039/501100011033 and by ``ESF Investing in your future''. This work has been funded through the European Research Council under grant ERC-2018-StG-802836
(AxScale project) as well as ERC-2017-AdG-788781 (IAXO+ project) and the Lise Meitner program of the Max Planck society. JD would like to acknowledge funding from the Royal Society under grant number URF/R1/211059.
BD and CC would like to thank Claudio Gatti and Alessio Rettaroli for very useful conversations regarding the sensitivity of the FLASH proposal and the sensitivity prospects to Dark Photons. We also thank V. Domcke for discussions about the sensitivity to HFGWs.

We sincerely thank the CERN cryolab team, in particular Torsten K\"ottig and Patricia Borges, for their advise and guidance to design and test a concept for a cryostat for the BabyIAXO magnet bore.

\end{document}